%% file: Driver.tex
\documentclass[reivew,number,sort&compress]{elsarticle}
\journal{Computer Methods in Applied Mechanics and Engineering}

\usepackage{graphics,graphicx}
\usepackage{tabularx}
\usepackage[parfill]{parskip}
\usepackage{amssymb,amsmath}
\usepackage{bm}
\usepackage{algorithm,algorithmic}
\usepackage{float}
\usepackage{units}
\usepackage{subfigure}
\usepackage{comment}

\newproof{proof}{Proof}
\newtheorem{corollary}{Corollary}
\newtheorem{proposition}{Proposition}

\graphicspath
{
    {./}
}

\newcommand{\mbs}[1]{\boldsymbol{#1}}

 \def\bB{{\mbs{B}}}

 \def\bQ{{\mbs{Q}}} \def\bR{{\mbs{R}}}

 \def\be{{\mbs{e}}} \def\fb{{\mbs{f}}}

 \def\bn{{\mbs{n}}} 
  
\def\bs{{\mbs{s}}} \def\bt{{\mbs{t}}} \def\bu{{\mbs{u}}}
  \def\bx{{\mbs{x}}}

\begin{document}

\begin{frontmatter}

\title{Data-driven computational mechanics}

\author[CIT]{T.~Kirchdoerfer}

\author[CIT]{M.~Ortiz\corref{fn:CIT}}
\ead{ortiz@caltech.edu}

\address[CIT]{Graduate Aerospace Laboratories,
California Institute of Technology \\ 1200 E. California Blvd., Pasadena, Ca, USA, 91125}

\cortext[fn:CIT]{1200 E. California Blvd., MC 105-50, Pasadena, Ca, USA, 91125}

\begin{abstract}
We develop a new computing paradigm, which we refer to as data-driven computing, according to which calculations are carried out directly from experimental material data and pertinent constraints and conservation laws, such as compatibility and equilibrium, thus bypassing the empirical material modeling step of conventional computing altogether. Data-driven solvers seek to assign to each material point the state from a prespecified data set that is closest to satisfying the conservation laws. Equivalently, data-driven solvers aim to find the state satisfying the conservation laws that is closest to the data set. The resulting data-driven problem thus consists of the minimization of a distance function to the data set in phase space subject to constraints introduced by the conservation laws. We motivate the data-driven paradigm and investigate the performance of data-driven solvers by means of two examples of application, namely, the static equilibrium of nonlinear three-dimensional trusses and linear elasticity. In these tests, the data-driven solvers exhibit good convergence properties both with respect to the number of data points and with regard to local data assignment. The variational structure of the data-driven problem also renders it amenable to analysis. We show that, as the data set approximates increasingly closely a classical material law in phase space, the data-driven solutions converge to the classical solution. We also illustrate the robustness of data-driven solvers with respect to spatial discretization. In particular, we show that the data-driven solutions of finite-element discretizations of linear elasticity converge jointly with respect to mesh size and approximation by the data set.
\end{abstract}

\begin{keyword}
data science \sep big data \sep approximation theory \sep scientific computing
\end{keyword}

\end{frontmatter}

\section{Introduction}

\input Introduction.tex

\section{Truss structures}\label{sec:Truss}

\input Trusses.tex

\section{Linear elasticity}\label{sec:LE}

\input LinearElasticity.tex

\section{Mathematical analysis of convergence}

\input Analysis.tex

\section{Summary and concluding remarks}

\input Summary.tex

\section*{Acknowledgments}

The support of Caltech's Center of Excellence on High-Rate Deformation Physics of Heterogeneous Materials, AFOSR Award FA9550-12-1-0091, is gratefully acknowledged.

\clearpage

\section*{Bibliography}

\bibliographystyle{elsarticle-num}
\bibliography{biblio}

\end{document}

%% file: Introduction.tex
Boundary-value problems in science and engineering typically combine two types of equations: i) {\sl Conservation laws}, which derive from universal principles such as conservation of momentum or energy and are, therefore, uncertainty-free; and ii) {\sl material laws}, formulated through physical modeling based on experimental observation, that are, therefore, empirical and uncertain. The prevailing classical computational paradigm has been to calibrate empirical material models using observational data and then use the calibrated material model in calculations. This process of modeling {\sl a fortiriori} adds error and uncertainty to the solutions, especially in systems with high-dimensional phase spaces and complex behavior. This modeling error and uncertainty arise from imperfect knowledge of the functional form of the material laws, the phase space in which they are defined, and from scatter and noise in the experimental data. Furthermore, often the models used to fit the data are {\sl ad hoc}, without a clear basis in physics or a mathematical criterion for their selection, and thus the process of modeling is mired in empiricism and arbitrariness. Indeed, the entire process of empirical material modeling, and model validation thereof, is open-ended and no rigorous mathematical theory exists to date that makes it precise and quantitative.

Previous work has been carried out with a view to incorporating observational data into boundary-value problem solution methodologies, but typically with the aim of parametric identification, or augmenting and automating, rather than replacing, the use and generation of material models. Material informatics uses database techniques to first identify parameters of correlation and then use machine-learning regression techniques \cite{Bishop:2006} to ultimately provide predictive quantitative models \cite{Rajan:2005}. Principal-component analysis provides methods of dimensional reduction that allow such modeling techniques to be applied \cite{Curtarolo:2003}.  These approaches have been extended to the generation of multi-scale modeling correlations between macroscopic and microscopic constitutive properties \cite{Breneman:2013,Kalidindi:2011,Krein:2012,Kalidindi:2015,Gupta:2015}.

These efforts, and others like them, may be understood as instances of {\sl Data Science}, the extraction of ‘knowledge’ from large volumes of unstructured data \cite{Agarwal:2014, Baesens:2014}. Data science often requires sorting through big-data sets and extracting ‘insights’ from these data. Data science uses data management, statistics and machine learning to derive mathematical models for subsequent use in decision making. Data Science currently influences primarily fields such as marketing, advertising, finance, social sciences, security, policy, medical informatics, whereas the full potential of Data Science as it relates to high-performance scientific computing is yet to be realized. Despite these limitations, reference to Data Science does effectively serve the purpose of bringing data and artificial intelligence considerations to the forefront.

In this work, we propose a new and different paradigm, which we refer to as {\sl data-driven computing}, consisting of formulating calculations {\sl directly} from experimental material data and pertinent essential constraints and conservation laws, thus bypassing the empirical material modeling step of conventional computing altogether. In this new computing paradigm, essential constraints and conservation laws such as compatibility and equilibrium remain unchanged, as do all the numerical schemes used in their discretization, such as finite elements, time-integrators, {\sl etc}. Such conservation laws confer mathematical structure to the calculations, and this mathematical structure carries over to the present data-driven paradigm. However, in sharp contrast to conventional computing, in data-driven computing the experimental material-data points are used directly in calculations {\sl in lieu} of an empirical material model. In this manner, material modeling empiricism, error and uncertainty are eliminated entirely and no loss of experimental information is incurred. Specifically, data-driven solvers seek to assign to each material point the state from a prespecified data set that is closest to satisfying the conservation laws. Equivalently, data-driven solvers aim to find the state satisfying the conservation laws that is closest to the data set. The resulting data-driven problem thus consists of the minimization of a distance function to the data set in phase space subject to the satisfaction of essential constraints and conservation laws.

We provide an efficient implementation of data-driven computing and demonstrate the practicality of the approach by means of two examples of application, namely, the static equilibrium of nonlinear three-dimensional trusses and linear elasticity. In these tests, the data-driven solvers exhibit good convergence properties both with respect to the number of data points and with regard to local data assignment. The variational structure of the data-driven problem also renders it amenable to analysis. We show that, as the data set approximates increasingly closely a classical material law in phase space, the data-driven solutions converge to the classical solutions. We also illustrate the robustness of data-driven solvers with respect to spatial discretization. In particular, we show that the data-driven solutions of finite-element discretizations of linear elasticity converge jointly with respect to mesh size and approximation by the data set. The mathematical analysis is also suggestive of a number of generalizations and extensions of the data-driven computing paradigm.

%% file: Trusses.tex
\begin{figure} [H]
\centering
\includegraphics[width=0.75\linewidth]{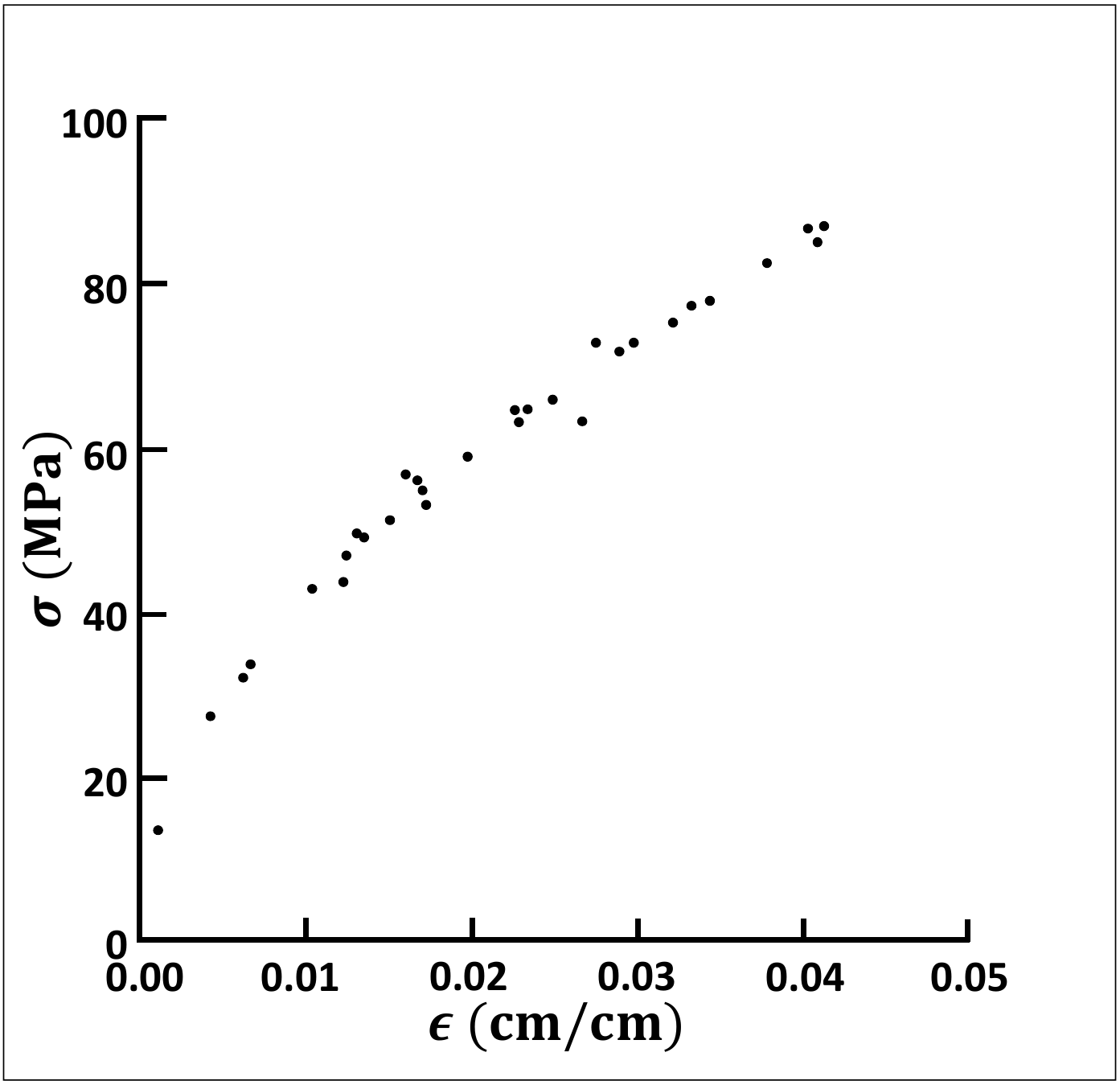}
\caption{Typical material data set for truss bar.}
\label{fig:Approx:Fig1a}
\end{figure}

We proceed to introduce and motivate the general approach with the aid of a simple non-linear elastic truss problem. Trusses are assemblies of articulated bars that deform in uniaxial tension or compression. Therefore, the material behavior of a bar is characterized by a particularly simple relation between uniaxial strain $\varepsilon$ and uniaxial stress $\sigma$. We refer to the space of pairs $(\varepsilon, \sigma)$ as {\sl phase space}. We assume that the behavior of the material of each bar $e=1,\dots,m$, where $m$ is the number of bars in the truss, is characterized by---possibly different---sets $E_e$ of pairs $(\varepsilon, \sigma)$, or {\sl local states}. For instance, each point in the data set may correspond to, e.~g., an experimental measurement, a subgrid multiscale calculation, or some other means of characterizing material behavior. A typical data set is notionally depicted in Fig.~\ref{fig:Approx:Fig1a}.

\subsection{Data-driven solver}

For a given material data set, the proposed data-driven solvers seek to assign to each bar $e=1,\dots,m$ of the truss the best possible local state ($\varepsilon_e, \sigma_e)$ from the corresponding data set $E_e$, while simultaneously satisfying compatibility and equilibrium. We understand optimality of the local state in terms of an appropriate figure of merit that penalizes distance to the data set in phase space. For definiteness, we consider local penalty functions of the type
\begin{equation}\label{eq:Truss:Fe}
    F_e(\varepsilon_e,\sigma_e)
    =
    \min_{(\varepsilon_e',\sigma_e') \in E_e}
    \Big(
        W_e(\varepsilon_e-\varepsilon_e')
        +
        W_e^*(\sigma_e-\sigma_e')
    \Big) ,
\end{equation}
for each bar $e=1,\dots,m$ in the truss, with
\begin{equation}
    W_e(\varepsilon_e)
    =
    \frac{1}{2} C_e \varepsilon_e^2
    \quad \text{and} \quad
    W_e^*(\sigma_e)
    =
    \frac{1}{2} \frac{\sigma_e^2}{C_e}
\end{equation}
and with the minimum taken over all local states $(\varepsilon_e',\sigma_e')$ in the local data set $E_e$. We may regard $W_e$ and $W_e^*$ are reference strain and complementary energy densities, respectively. We emphasize that the functions $W_e$ and $W_e^*$ are introduced as part of the numerical scheme and need not represent any actual material behavior. In particular, the constant $C_e$ is also numerical in nature and does not represent a material property.

Given a global state consisting of the collection of the local states $(\varepsilon_e, \sigma_e)$ of each one of its bars, the combined penalty function
\begin{equation}\label{eq:Truss:F}
    F
    =
    \sum_{e=1}^m w_e
    F_e(\varepsilon_e,\sigma_e)
\end{equation}
simultaneously penalizes all local departures of the local states of the bars from their corresponding data sets. Here and subsequently, $w_e = A_e L_e$ denotes the volume of truss member $e$, with $A_e$ its cross-sectional area and $L_e$ its length. The aim of the data-driven solver is to minimize $F$ with respect to the global state $\{(\varepsilon,\sigma)\}$ subject to equilibrium and compatibility constraints. These considerations lead to the constrained minimization problem
\begin{subequations}
\begin{align}
    &
    \text{Minimize:}\quad
    \sum_{e=1}^m w_e  F_e(\varepsilon_e,\sigma_e) ,
    \\ &
    \text{subject to:}\quad
    \varepsilon_e
    =
    \sum_{i=1}^n B_{ei} u_i
    \quad \text{and} \quad
    \sum_{e=1}^m w_e  B_{ei} \sigma_e = f_i ,
\end{align}
\end{subequations}
where $\{u_i,\ i=1,\dots,n\}$ is the array of displacement degrees of freedom, $\{f_i,\ i=1,\dots,n\}$ is the array of applied forces and the coefficients $B_{ei}$ encode the connectivity and geometry of the truss.

The compatibility constraint can be enforced simply by expressing the strains in terms of displacements. The equilibrium constraint can be enforced by means of Lagrange multipliers, leading to the stationary problem
\begin{equation}
    \delta
    \left(
    \sum_{e=1}^m w_e  F_e(\sum_{i=1}^n B_{ei} u_i,\sigma_e)
    -
    \sum_{i=1}^N \Big(\sum_{e=1}^m w_e  B_{ei} \sigma_e - f_i \Big) \eta_i
    \right)
    =
    0 .
\end{equation}
Taking all possible variations, we obtain
\begin{subequations}\label{eq:Truss:EL}
\begin{align}
    &
    \delta u_i \Rightarrow
    \sum_{e=1}^m w_e
        C_e
        \Big(
            \sum_{j=1}^n B_{ej} u_j-\varepsilon_e^*
        \Big)
        B_{ei}
    =
    0 ,
    \\ &
    \delta \sigma_e \Rightarrow
    \frac{1}{C_e}(\sigma_e-\sigma_e^*)
    =
    \sum_{i=1}^n B_{ei} \eta_i ,
    \label{eq:Truss:EL2}
    \\ &
    \delta \eta_i \Rightarrow
    \sum_{e=1}^m w_e   B_{ei} \sigma_e
    =
    f_i ,
\end{align}
\end{subequations}
where $(\varepsilon_e^*, \sigma_e^*)$ denote (unknown) optimal data points for each of the bars, i.~e., data points such that
\begin{equation}
    F_e(\sum_{i=1}^n B_{ei} u_i,\sigma_e)
    =
    W_e(\sum_{i=1}^n B_{ei} u_i-\varepsilon_e^*)
    +
    W_e^*(\sigma_e-\sigma_e^*)
\end{equation}
or
\begin{equation}
    W_e(\sum_{i=1}^n B_{ei} u_i-\varepsilon_e^*)
    +
    W_e^*(\sigma_e-\sigma_e^*)
    \leq
    W_e(\sum_{i=1}^n B_{ei} u_i-\varepsilon_e')
    +
    W_e^*(\sigma_e-\sigma_e')
\end{equation}
for all data points $(\varepsilon_e', \sigma_e')$ in the local data set $E_e$. Once all optimal data points are determined, eqs.~(\ref{eq:Truss:EL}) define a system of linear equations for the nodal displacements, the local stresses and the Lagrange multipliers. A straightforward manipulation of these equations renders them in the equivalent form
\begin{subequations}\label{eq:Truss:ELred}
\begin{align}
    &
    \sum_{j=1}^n
        \left( \sum_{e=1}^m w_e  C_e B_{ej} B_{ei} \right) u_j
    =
    \sum_{e=1}^m w_e
        C_e \varepsilon_e^* B_{ei} ,
    \label{eq:Truss:ELred1}
    \\ &
    \sum_{j=1}^n
        \left(\sum_{e=1}^m w_e   C_e B_{ei}B_{ej} \right) \eta_j
    =
    f_i - \sum_{e=1}^m w_e   B_{ei} \sigma_e^* .
    \label{eq:Truss:ELred2}
\end{align}
\end{subequations}
We recognize in these equations two standard linear-elastic truss-equilibrium problems with identical stiffness matrix corresponding to the reference linear truss defined by $W_e$ and $W_e^*$, $e=1,\dots,m$. The displacement problem (\ref{eq:Truss:ELred1}) is driven by the optimal local strains, whereas the Lagrange multiplier problem (\ref{eq:Truss:ELred2}) is driven by the out-of-balance forces attendant to the optimal local stresses.

It remains to determine the optimal local data points, i.~e., the stress and strain pairs $(\varepsilon_e^*, \sigma_e^*)$ in the local data sets $E_e$ that result in the closest possible satisfaction of compatibility and equilibrium. The determination of the optimal local data points can be effected iteratively. Initially, all bars in the truss are assigned random points $(\varepsilon_e^{*(0)}, \sigma_e^{*(0)})$ from the corresponding local data sets $E_e$. The displacements $u_i^{(0)}$ and Lagrange multipliers $\eta_i^{(0)}$ are then computed by solving (\ref{eq:Truss:ELred}) and the stresses $\sigma_e^{(0)}$ are evaluated from (\ref{eq:Truss:EL2}). The next local data assignment is then effected by determining, for every member in the truss, the data points $(\varepsilon_e^{*(1)}, \sigma_e^{*(1)})$ in $E_e$ that are optimal with respect to the local state $(\varepsilon_e^{(0)}, \sigma_e^{(0)})$, i.~e., such that
\begin{equation}\label{eq:Truss:loc}
    W_e(\varepsilon_e^{(0)}-\varepsilon_e^{*(1)})
    +
    W_e^*(\sigma_e^{(0)}-\sigma_e^{*(1)})
    \leq
    W_e(\varepsilon_e^{(0)}-\varepsilon_e')
    +
    W_e^*(\sigma_e^{(0)}-\sigma_e') ,
\end{equation}
for all data points $(\varepsilon_e', \sigma_e')$ in the local data set $E_e$. This operation entails simple local searches in phase space. The iteration then proceeds by recursion and terminates when the local data assignments effect no change. A detailed flowchart of the data-driven solver is listed in Algorithm~\ref{alg:Solver}.

\begin{algorithm}[H]
\caption{Data-driven solver}
\label{alg:Solver}
\begin{algorithmic}

\REQUIRE Local data sets $E_e$, $B_e$-matrices, $e=1,\dots,m$. Applied loads $f_i$, $i=1,\dots,n$.
\STATE i) Set $k=0$. Initial local data assignment:
\FORALL {$e=1,\dots,m$}
\STATE Choose $(\varepsilon_e^{*(0)}, \sigma_e^{*(0)})$ randomly from $E_e$
\ENDFOR
\STATE ii) Solve:
\begin{subequations}\label{eq:Truss:alg:EL}
\begin{align}
    &
    \sum_{j=1}^n
    \left( \sum_{e=1}^m w_e  C_e B_{ej} B_{ei} \right) u_j^{(k)}
    =
    \sum_{e=1}^m w_e
    C_e \varepsilon_e^{*(k)} B_{ei} ,
    \label{eq:Truss:alg:EL1}
    \\ &
    \sum_{j=1}^n
    \left(\sum_{e=1}^m w_e   C_e B_{ei}B_{ej} \right) \eta_j^{(k)}
    =
    f_i - \sum_{e=1}^m w_e   B_{ei} \sigma_e^{*(k)} ,
    \label{eq:Truss:alg:EL2}
\end{align}
\end{subequations}
for $u_i^{(k)}$ and $\eta_i^{(k)}$, $i=1,\dots,n$.
\STATE iii) Compute local states:
\FORALL {$e=1,\dots,m$}
\STATE
\begin{equation}\label{eq:Truss:alg:Loc}
    \varepsilon_e^{(k)}
    =
    \sum_{i=1}^n B_{ei} u_i^{(k)} ,
    \qquad
    \sigma_e^{(k)}
    =
    \sigma_e^{*(k)} + C_e \sum_{i=1}^n B_{ei} \eta_i^{(k)}
\end{equation}
\ENDFOR
\STATE iv) Local state assignment:
\FORALL {$e=1,\dots,m$}
\STATE Choose $(\varepsilon_e^{*(k+1)}, \sigma_e^{*(k+1)})$ closest to $(\varepsilon_e^{(k)}, \sigma_e^{(k)})$ in $E_e$.
\ENDFOR
\STATE v) Test for convergence:
\IF{$(\varepsilon_e^{*(k+1)}, \sigma_e^{*(k+1)})$ $=$ $(\varepsilon_e^{*(k)}, \sigma_e^{*(k)})$ for all $e=1,\dots,m$,}
\STATE v.a) $u_i = u_i^{(k)}$, $i=1,\dots,n$.
\STATE v.b) $(\varepsilon_e, \sigma_e)$ $=$ $(\varepsilon_e^{(k)}, \sigma_e^{(k)})$, $e=1,\dots,m$.
\STATE v.c) {\bf exit}.
\ELSE
\STATE $k \leftarrow k+1$, {\bf goto} (ii).
\ENDIF
\end{algorithmic}
\end{algorithm}

\begin{figure} [H]
\centering
\includegraphics[width=0.65\linewidth]{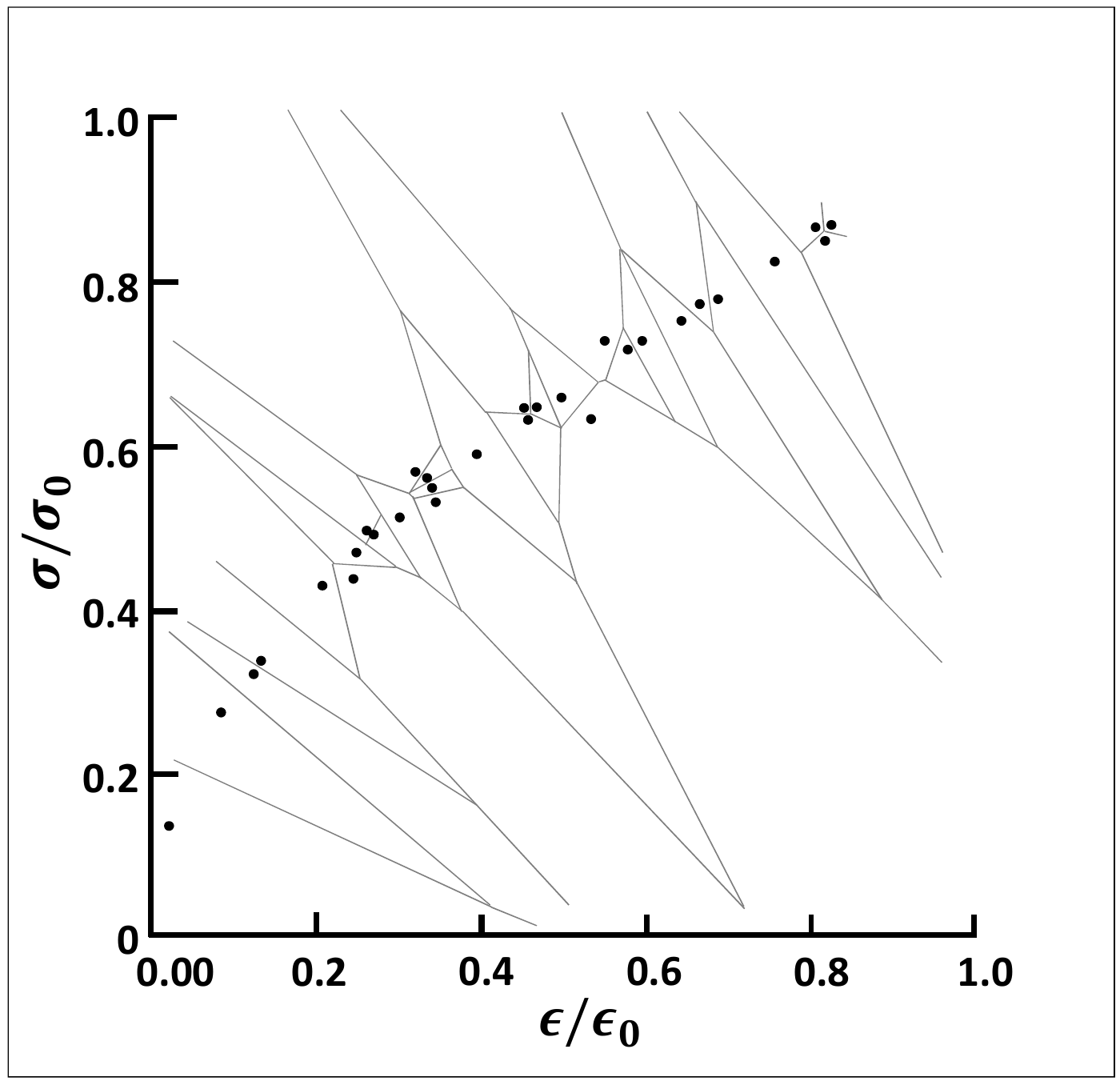}
\caption{Voronoi tessellation of a data set.}
\label{fig:Approx:Fig1b}
\end{figure}

The geometry of the local data assignment (\ref{eq:Truss:loc}) is illustrated in Fig.~\ref{fig:Approx:Fig1b}. Thus, given a trial local state $(\varepsilon_e^{(k)}, \sigma_e^{(k)})$ of bar $e$, corresponding to the $k$th iteration of the solver, the next data point $(\varepsilon_e^{*(k+1)}, \sigma_e^{*(k+1)})$ assigned to the bar is the point in $E_e$ that is closest to $(\varepsilon_e^{(k)}, \sigma_e^{(k)})$ in the norm
\begin{equation}\label{eq:Truss:norm}
    \| (\varepsilon_e, \sigma_e) \|_e
    =
    \Big(
        W_e(\varepsilon_e)
        +
        W_e^*(\sigma_e)
    \Big)^{1/2} .
\end{equation}
This is precisely the data point $(\varepsilon_e^{*(k+1)}, \sigma_e^{*(k+1)})$ in $E_e$ whose Voronoi cell contains $(\varepsilon_e^{(k)}, \sigma_e^{(k)})$. Thus, the penalty function (\ref{eq:Truss:Fe}) or, equivalently, the norm (\ref{eq:Truss:norm}) divides the phase space into cells according to the Voronoi tessellation of $E_e$. Each cell in that tessellation may be regarded as the 'domain of influence' of the corresponding data point. The local state assignment then simply assigns material points according to their domain of influence and the iteration terminates when the local states of all bars lie within the domain of influence of the corresponding data points assigned to the bars.

\subsection{Numerical analysis of convergence}

A central question to be ascertained concerns the convergence of data-driven solvers with respect to the data set. Specifically, suppose that the materials in the truss obey a well-defined constitutive law in the form of a graph, or stress-strain curve, in $(\varepsilon, \sigma)$-phase space. Then, we expect the data-driven solutions to converge to the classical solution when the data sets approximate the stress-strain curve increasingly closely, in some appropriate sense to be made precise subsequently.

\begin{figure} [H]
\centering
  \includegraphics[width=0.75\linewidth]{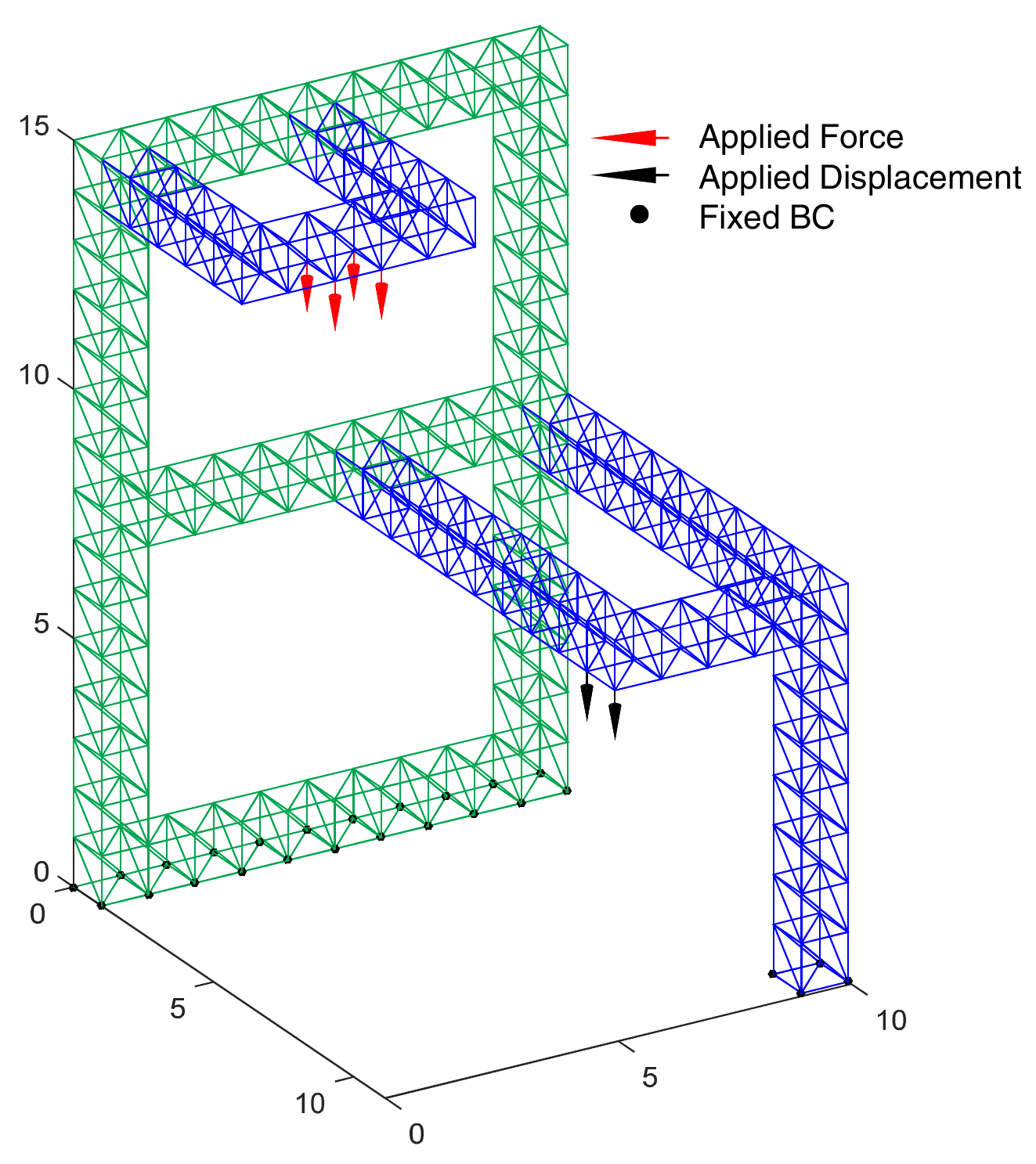}
\caption{Model problem geometry with boundary conditions.}
\label{fig:trussgeom}
\end{figure}

\begin{figure} [H]
\centering
  \includegraphics[width=0.75\linewidth]{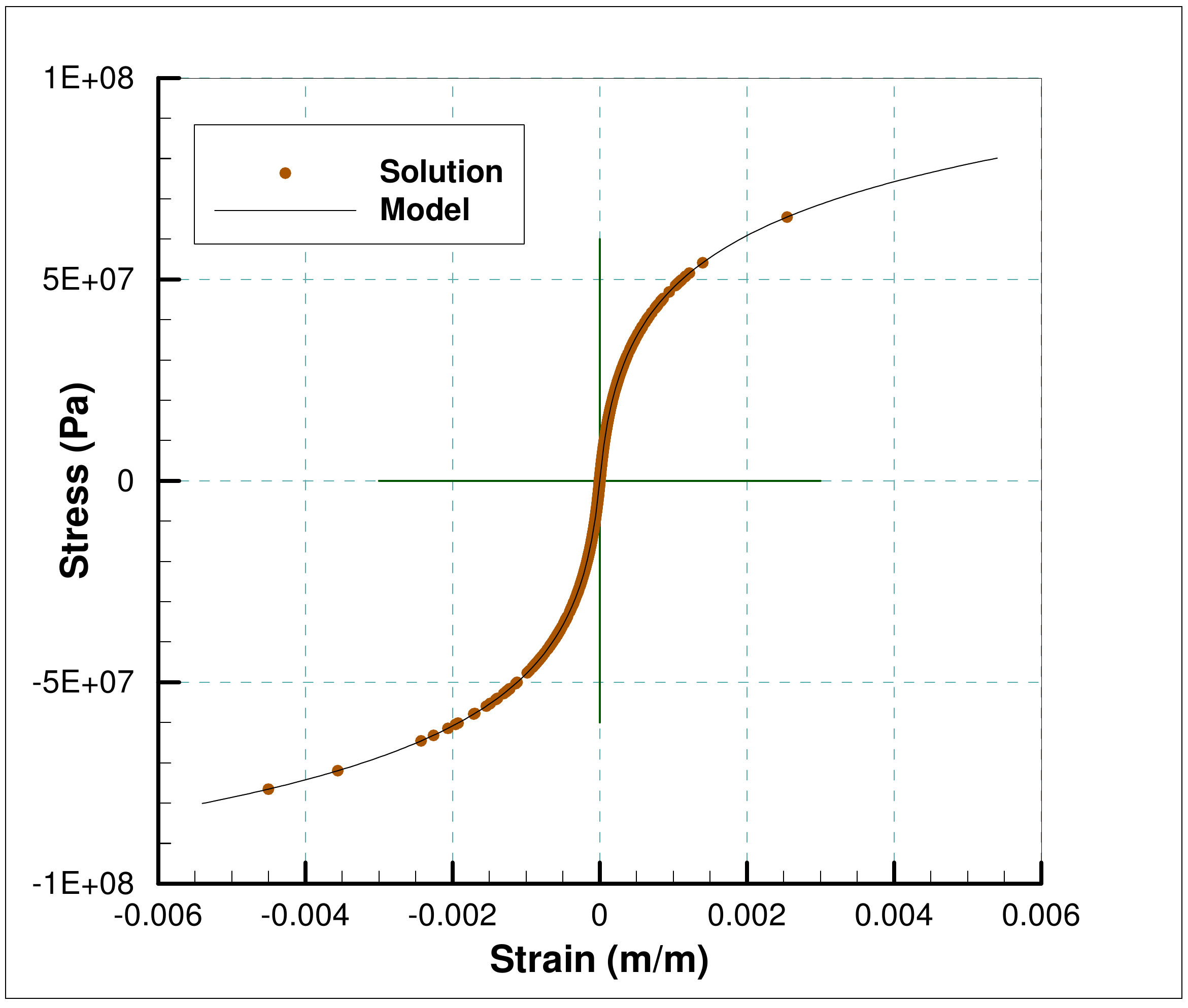}
\caption{Material model with reference solution values superimposed.}
\label{fig:trussmat}
\end{figure}

In this section, we exhibit this convergence property in a specific example of application. Fig.~\ref{fig:trussgeom} shows the geometry, boundary conditions and applied loads on a truss containing 1,048 degrees of freedom. The truss undergoes small deformations and the material in all bars obeys the nonlinear non-linear elastic law shown in Fig.~\ref{fig:trussmat}. A Newton-Raphson solver is used to calculate the reference solution. The reference solution values thus obtained are plotted on the constitutive stress-strain curve to exhibit the extent of non-linearity and the range of local states covered by the solution.

Suppose that, in actual practice, the stress-strain curve in Fig.~\ref{fig:trussmat} is not known exactly but, instead, sampled by means of a finite collection of points, or data sets. We begin by considering a sequence $(E_k)$ of increasingly fine data sets consisting of points on the stress-strain curve at uniform distances $\rho_k \downarrow 0$, with distance defined in the sense of the norm (\ref{eq:Truss:norm}).

\begin{figure} [H]
\centering
  \includegraphics[width=0.75\linewidth]{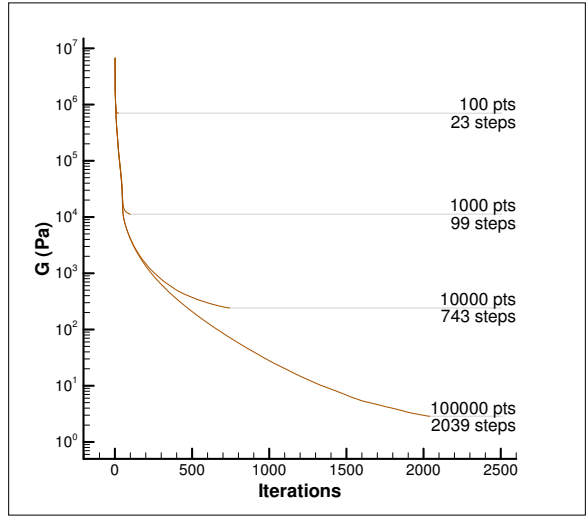}
\caption{Convergence of the local data-assignment iteration.}
\label{fig:stepconv}
\end{figure}

The convergence of the local data assignment iteration is shown in Fig.~\ref{fig:stepconv} for data sets of sizes $10^2$, $10^3$, $10^4$, $10^5$. In all cases, the initial local data assignment is random and convergence is monitored in terms of the penalty function $F$, eq.~(\ref{eq:Truss:F}). We note that the problem of assigning data points optimally to each bar of the truss is of combinatorial complexity. Therefore, it is remarkable that the local data assignment iteration converges after a relatively small number of steps. As expected, the number of iterations to convergence increases with the size of the data set. However, it bears emphasis that each local data assignment iteration entails a linear solve corresponding to the linear comparison truss. The matrix of the system of equations, or stiffness matrix, can be factorized once and for all at the start of the iteration and subsequent iterations require inexpensive back-substitutes only.

\begin{figure}[H]
  \centering
  \includegraphics[width=0.75\linewidth]{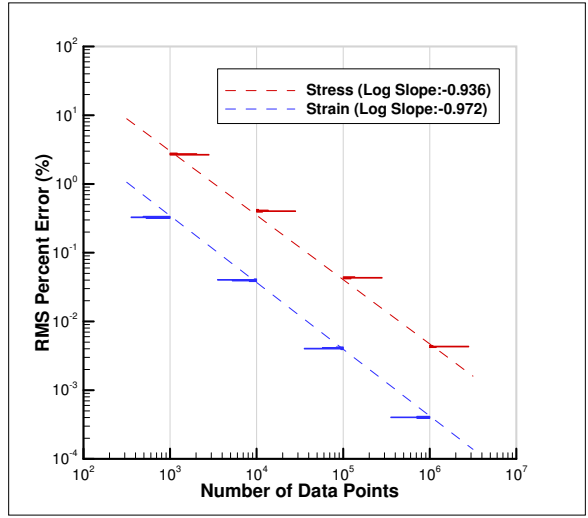}
  \caption{Convergence of strain and stress root-mean-square errors with number of sampling points. Histograms correspond to $30$ different initial random assignments of data points to the truss members.}
  \label{fig:NoNoiseConverge}
\end{figure}

Next, we turn to the question of convergence with respect to the number of data points. For definiteness, we monitor the convergence of the resulting sequence of data-driven solutions to the reference solution in the sense of the normalized percent root-mean-square stress and strain errors
\begin{subequations}\label{eq:rmsdef}
  \begin{align}
    \varepsilon_{(\%{\rm RMS})}
    =
    \frac{1}{\varepsilon^{\rm ref}_{\rm max}}
    \left(
        \frac{\sum_{e=1}^m w_e  {(\varepsilon_e-\varepsilon^{\rm ref}_e)^2}}{m}
    \right)^{1/2} ,
    \\
    \sigma_{(\%{\rm RMS})}
    =
    \frac{1}{\sigma^{\rm ref}_{\rm max}}
    \left(
        \frac{\sum_{e=1}^m w_e {(\sigma_e-\sigma^{\rm ref}_e)^2}}{m}
    \right)^{1/2} ,
  \end{align}
\end{subequations}
respectively, where $(\varepsilon^{\rm ref}_e, \sigma^{\rm ref}_e)$, $e = 1,\dots,m$ are the strains and stresses corresponding to the reference solution and $(\varepsilon^{\rm ref}_{\rm max}, \sigma^{\rm ref}_{\rm max})$ are the corresponding maximum values.

Fig.~\ref{fig:NoNoiseConverge} shows convergence plots of the strain and stress root-mean-square errors with number of sampling points. As may be observed from the figure, the convergence is close to linear in both strains and stresses, which verifies the convergence of the method as the data set approaches the exact model. We recall that the data assignment algorithm~\ref{alg:Solver} starts by randomly assigning data points to the truss members. Evidently, the subsequent iteration depends on this initial choice. In order to demonstrate insensitivity to such initialization, convergence plots for $30$ initial random assignments are shown in Fig.~\ref{fig:NoNoiseConverge} and the resulting errors are binned into histograms. The tightness of these histograms verifies the robustness of the iteration with respect to the initial data point selection.

\begin{figure} [H]
\centering
  \includegraphics[width=0.75\linewidth]{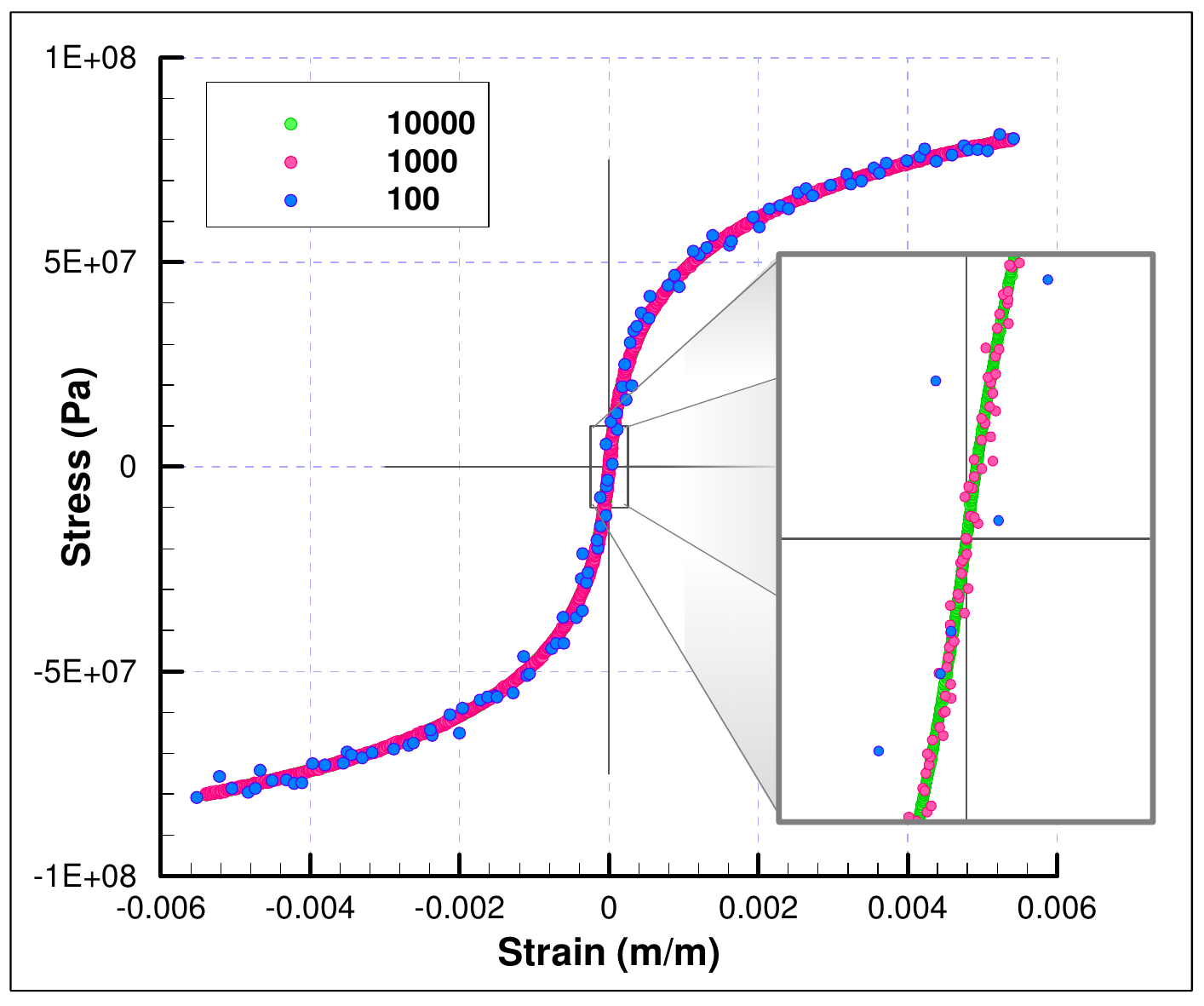}
\caption{Typical data set with Gaussian random noise.}
\label{fig:Noisy}
\end{figure}

\begin{figure}[H]
  \centering
  \includegraphics[width=0.75\linewidth]{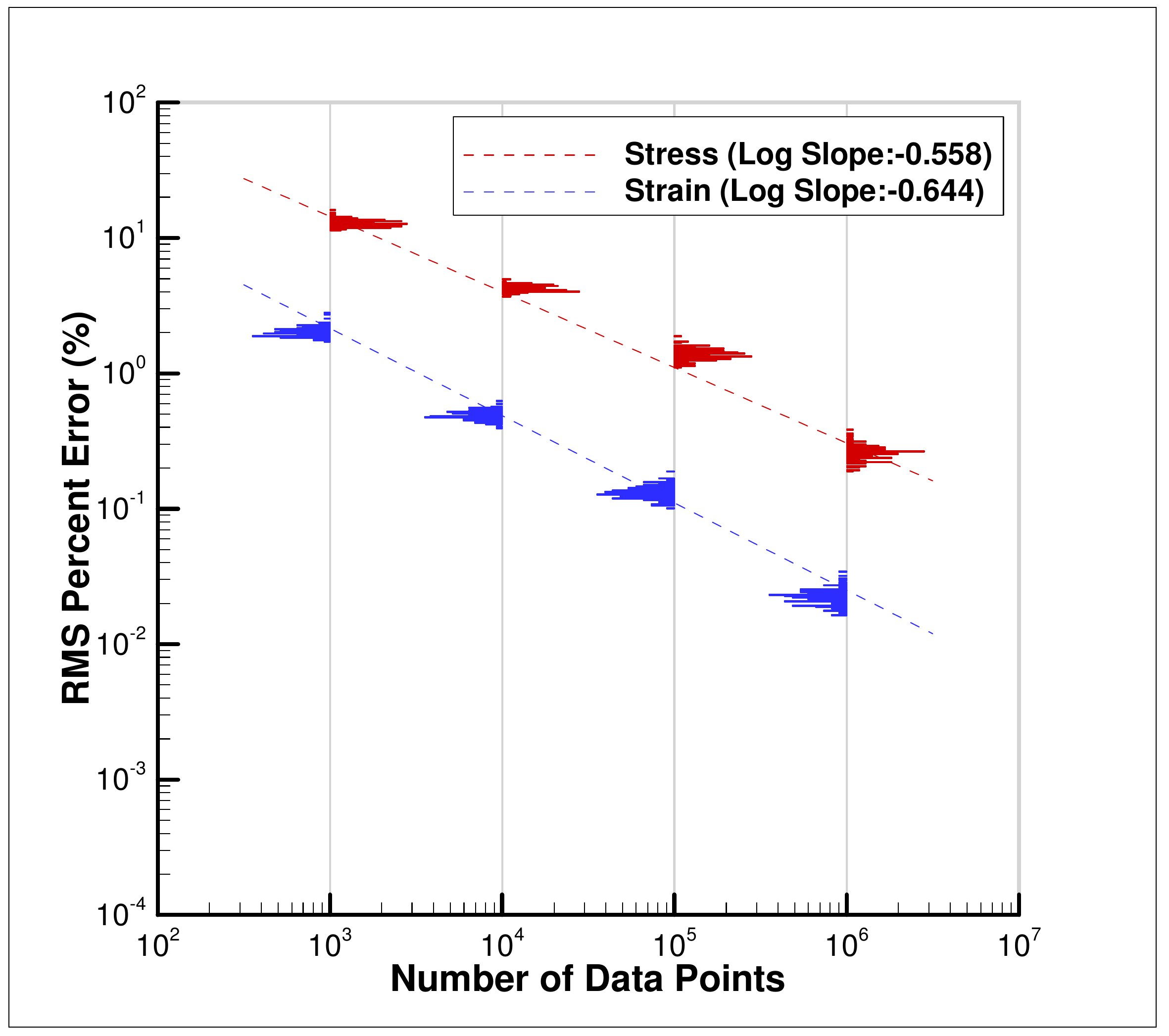}
  \caption{Convergence of strain and stress root-mean-square errors with number of sampling points and data sets with Gaussian noise. Histograms correspond to $100$ data sets.}
  \label{fig:NoiseConverge}
\end{figure}

Next, we revisit the question of convergence with respect to the number of data points when the data set is noisy, i.~e., when it does not sample the limit stress-strain curve but is offset from the curve with some probability. In this case, the data sets converge to the exact stress-strain curve as sets, in a manner to be made precise subsequently. In calculations we specifically begin by sampling the limit stress-stain curve at uniform distances $\rho_k \downarrow 0$, as in the preceding test cases, and subsquently add Gaussian noise to the data points of variance $\rho_k$. A typical data set is shown in Fig.~\ref{fig:Noisy} by way of illustration. Convergence plots corresponding to $100$ data sets are shown in Fig.~\ref{fig:NoiseConverge}. As may be seen from the figure, convergence is achieved with increasing number of points, albeit the convergence rate of roughly $1/2$ is lower than the convergence rate in the case of noiseless data.

\begin{figure} [H]
\centering
  \includegraphics[width=0.75\linewidth]{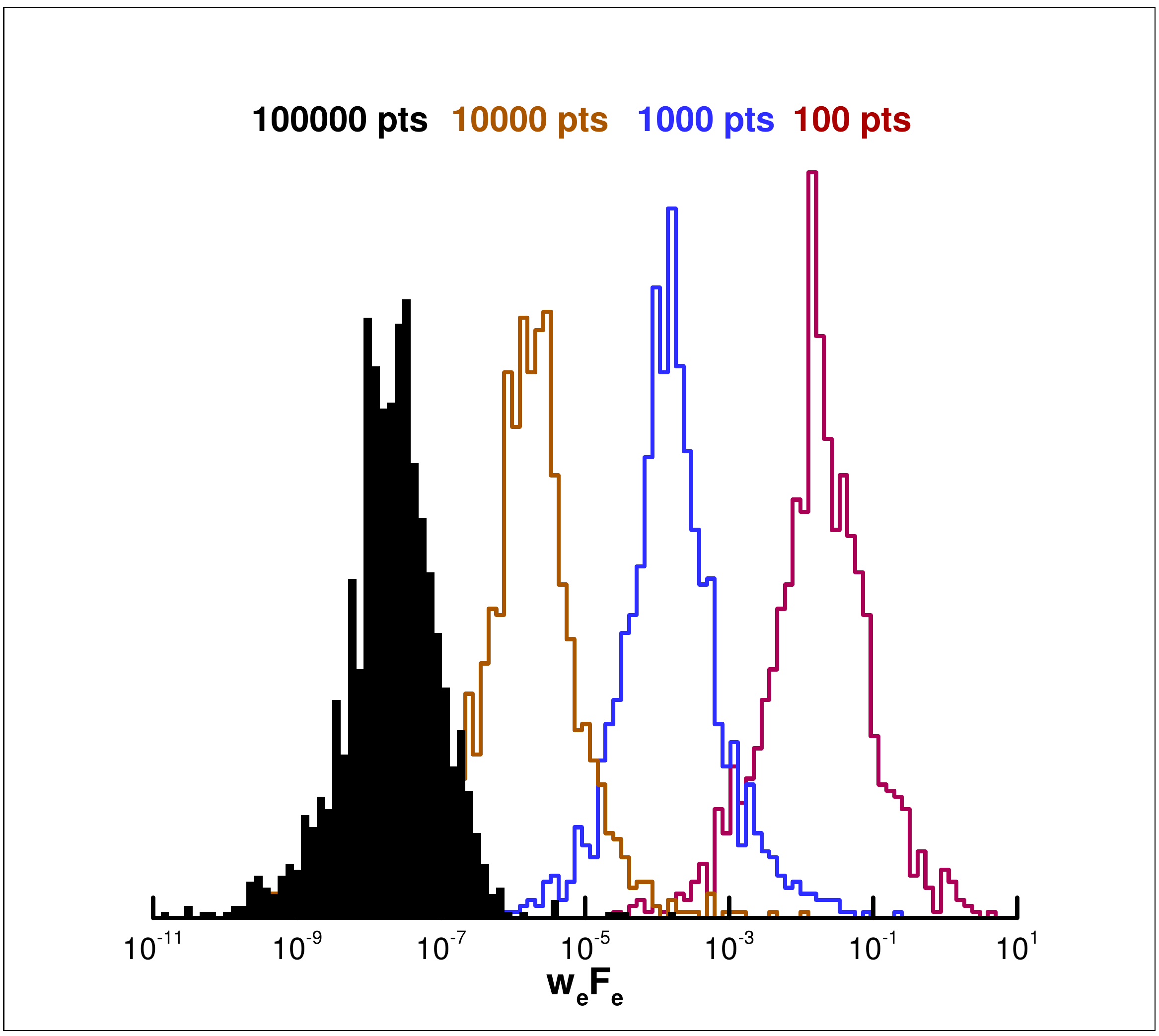}
\caption{Distribution of values of local penalty functions $F_e(\varepsilon, \sigma)$ for converged data-driven solution.}
\label{fig:Fdist}
\end{figure}

Finally, we examine the question of sample quality, i.~e., the ability of a given data set to sample closely all the local states covered by the solution. Fig.~\ref{fig:Fdist} shows the distribution of the values of the local penalty function $F_e$, eq.~(\ref{eq:Truss:Fe}) corresponding to data sets of sizes $10^2$, $10^3$, $10^4$, $10^5$. We recall that the value of the function $F_e$ provides a measure of the distance of the local state $(\varepsilon_e, \sigma_e)$ to the data set. As may be seen from the figure, $F_e$ tends to decrease with the number of sampling points, as expected. However, for every data-set size there remains a certain spread in the values of $F_e$, indicating that the states of certain truss members are better sampled by the data set than others. Specifically, truss members for which no data point lies close to their states result in high values of $F_e$, indicative of poor coverage by the data set. This analysis of the local values $F_e$ of the penalty function suggests a criterion for improving data sets adaptively so as to improve their quality vis a vis a particular application. Evidently, the optimal strategy is to target for further testing the region of phase space corresponding to the truss members with highest values of $F_e$. In particular, outliers, or truss members with states lying far from the data set, are targeted for further testing. In this manner, the data set is adaptively expanded so as to provide the best possible coverage of the distribution of local states corresponding to a particular application.

%% file: LinearElasticity.tex
As a second motivational example of application of the data-driven paradigm, we consider three-dimensional linear elasticity. In this case, the local phase space of the material consists of pairs $(\mbs{\epsilon},\mbs{\sigma})$ of strain and stress, respectively. Since both stresses and strains are symmetric tensors, it follows that the corresponding phase space is twelve-dimensional. This dimensionality is high enough to start raising questions regarding material sampling and material-data coverage of the relevant region of phase space. An additional issue that is raised by linear elasticity concerns the infinite-dimensional character of the solution space. Thus, even if the problem is rendered finite-dimensional by recourse to spatial discretization, the question of convergence with respect to mesh size must necessarily be elucidated within an appropriate functional framework. In this section, we extend the truss data-driven solver to linear elasticity and address the issue of data sampling in high dimensions by exploiting material and geometrical symmetry in the problem. Finally, we address the question of convergence of the finite-element discretized data-driven solver with respect to mesh size.

\subsection{Data-driven solver}

We consider a finite-element model of a nonlinear-elastic solid in the linearized kinematics approximation. The material behavior of a solid is characterized by a relation between the strain tensor $\mbs{\epsilon}$ and the stress tensor $\mbs{\sigma}$. We refer to the space of pairs $(\mbs{\epsilon}, \mbs{\sigma})$ as {\sl phase space}. We assume that the behavior of the material or integration points in the model is characterized by---possibly different---sets $E_e$ of pairs $(\mbs{\epsilon}, \mbs{\sigma})$, or {\sl local states}, where $e=1,\dots,m$ labels the material points and $m$ is the number of material points in the finite-element model.

We consider local penalty functions of the type
\begin{equation}
    F_e(\mbs{\epsilon}_e,\mbs{\sigma}_e)
    =
    \min_{(\mbs{\epsilon}_e',\mbs{\sigma}_e') \in E_e}
    \Big(
        W_e(\mbs{\epsilon}_e-\mbs{\epsilon}_e')
        +
        W_e^*(\mbs{\sigma}_e-\mbs{\sigma}_e')
    \Big) ,
\end{equation}
for each integration point $e=1,\dots,m$ in the solid, with
\begin{subequations}\label{eq:ury933}
\begin{align}
    &
    W_e(\mbs{\epsilon}_e)
    =
    \frac{1}{2}
    \lambda
    ({\rm tr} \mbs{\epsilon}_e)^2
    +
    \mu
    \mbs{\epsilon}_e \cdot \mbs{\epsilon}_e
    \equiv
    \mathbb{C}_e\mbs{\epsilon}_e \cdot \mbs{\epsilon}_e ,
    \\ &
    W_e^*(\mbs{\sigma}_e)
    =
    \frac{1}{4\mu}\mbs{\sigma}_e\cdot\mbs{\sigma}_e
    -
    \frac{1}{4\mu}\frac{\lambda}{3\lambda+2\mu}
    ({\rm tr} \mbs{\sigma}_e)^2
    \equiv
    \mathbb{C}_e^{-1}\mbs{\sigma}_e\cdot\mbs{\sigma}_e ,
\end{align}
\end{subequations}
with the minimum taken over all local states $(\mbs{\epsilon}_e',
\mbs{\sigma}_e')$ in the local data set $E_e$. We may regard $W_e$ and $W_e^*$ are reference strain and complementary energy densities, respectively.

Given a global state consisting of a collection of local states $(\mbs{\epsilon}_e, \mbs{\sigma}_e)$ at each material point, we define a global penalty function as
\begin{equation}
    F
    =
    \sum_{e=1}^m
    w_e
    F_e(\mbs{\epsilon}_e,\mbs{\sigma}_e) ,
\end{equation}
$w_e$ are quadrature or integration weights. This function penalizes jointly all departures of local states from their corresponding data sets. The data-driven problem is to minimize $F$ with respect to the global state $\{(\mbs{\epsilon}, \mbs{\sigma})\}$ subject to equilibrium and compatibility constraints, namely,
\begin{subequations}
\begin{align}
    &
    \text{Minimize:}\quad
    \sum_{e=1}^m w_e F_e(\mbs{\epsilon}_e,\mbs{\sigma}_e) ,
    \\ &
    \text{subject to:}\quad
    \mbs{\epsilon}_e
    =
    \sum_{a=1}^n \bB_{ea} \bu_a
    \quad \text{and} \quad
    \sum_{e=1}^m w_e \bB_{ea}^T \mbs{\sigma}_e = \fb_a ,
\end{align}
\end{subequations}
where $\{\bu_a,\ a=1,\dots,n\}$ is the array of nodal displacements, $\{\fb_a,\ a=1,\dots,n\}$ is the array of applied nodal forces, $n$ is the number of nodes and the coefficients $\bB_{ea}$ encode the connectivity and geometry of the finite-element mesh.

As in the data-driven truss problem, the compatibility constraint can be enforced simply by expressing the strains in terms of displacements. The equilibrium constraint can in turn be enforced by means of Lagrange multipliers, resulting in the stationary problem
\begin{equation}
    \delta
    \left(
    \sum_{e=1}^m w_e F_e(\sum_{a=1}^n \bB_{ea} \bu_a, \mbs{\sigma}_e)
    -
    \sum_{a=1}^N \Big(\sum_{e=1}^m w_e \bB_{ea}^T \mbs{\sigma}_e - \fb_a \Big) \mbs{\eta}_a
    \right)
    =
    0 .
\end{equation}
Taking all possible variations, we obtain the system of Euler-Lagrange equations
\begin{subequations}\label{eq:jp88tm}
\begin{align}
    &
    \delta \bu_a \Rightarrow
    \sum_{e=1}^m
        w_e
        \bB_{ea}^T
        \mathbb{C}_e
        \Big(
            \sum_{b=1}^n \bB_{eb} \bu_b-\mbs{\epsilon}_e^*
        \Big)
    =
    0 ,
    \\ &
    \delta \mbs{\sigma}_e \Rightarrow
    \mathbb{C}_e^{-1}(\mbs{\sigma}_e-\mbs{\sigma}_e^*)
    =
    \sum_{a=1}^n \bB_{ea} \mbs{\eta}_a ,
    \\ &
    \delta \mbs{\eta}_a \Rightarrow
    \sum_{e=1}^m  w_e \bB_{ea}^T \mbs{\sigma}_e
    =
    \fb_a ,
\end{align}
\end{subequations}
where $(\mbs{\epsilon}_e^*, \mbs{\sigma}_e^*)$ denote the unknown optimal data points at material point $e$, i.~e., the data point such that
\begin{equation}
    F_e(\sum_{a=1}^n \bB_{ea} \bu_a,\mbs{\sigma}_e)
    =
    W_e(\sum_{a=1}^n \bB_{ea} \bu_a-\mbs{\epsilon}_e^*)
    +
    W_e^*(\mbs{\sigma}_e-\mbs{\sigma}_e^*) ,
\end{equation}
or
\begin{equation}
    W_e(\sum_{a=1}^n \bB_{ea} \bu_a-\mbs{\epsilon}_e^*)
    +
    W_e^*(\mbs{\sigma}_e-\mbs{\sigma}_e^*)
    \leq
    W_e(\sum_{a=1}^n \bB_{ea} \bu_a-\mbs{\epsilon}_e')
    +
    W_e^*(\mbs{\sigma}_e-\mbs{\sigma}_e') ,
\end{equation}
for all data points $(\mbs{\epsilon}_e', \mbs{\sigma}_e')$ in the local data set $E_e$. Once all optimal data points are determined, eqs.~(\ref{eq:jp88tm}) define a system of linear equations for the nodal displacements, the local stresses and the Lagrange multipliers. As in the data-driven truss problem, these equations can be rendered in the equivalent form
\begin{subequations}
\begin{align}
    &
    \sum_{b=1}^n
        \left( \sum_{e=1}^m w_e
            \bB_{ea}^T \mathbb{C}_e \bB_{eb} \right) \bu_b
    =
    \sum_{e=1}^m w_e
        \bB_{ea}^T \mathbb{C}_e \mbs{\epsilon}_e^*  ,
    \label{eq:2ttahe}
    \\ &
    \sum_{b=1}^n
        \left(\sum_{e=1}^m w_e
            \bB_{ea}^T \mathbb{C}_e \bB_{eb} \right) \mbs{\eta}_b
    =
    \fb_a - \sum_{e=1}^m  w_e \bB_{ea}^T \mbs{\sigma}_e^* .
    \label{eq:5dskvd}
\end{align}
\end{subequations}
Here we recognize two standard linear-elastic equilibrium problems with identical stiffness matrix corresponding to the reference linear solid defined by $W_e$ and $W_e^*$, $e=1,\dots,m$. The displacement problem (\ref{eq:2ttahe}) is driven by the optimal local strains, whereas the Lagrange multiplier problem (\ref{eq:5dskvd}) is driven by the out-of-balance forces attendant to the optimal local stresses.

\subsection{Using material symmetries to reduce data sets}
\label{Sec:Isotropy}

Phase-space sampling requirements can be reduced if {\sl a priori} knowledge of material behavior is available. In particular, material symmetry can be effectively exploited for purposes of reducing material sampling requirements. A simple and commonly encountered example of material symmetry is isotropy. For a three-dimensional isotropic material in the linearized kinematics approximation, if $(\mbs{\epsilon}_e, \mbs{\sigma}_e)$ is a material data point, then so are $({\bR_e}^T \mbs{\epsilon}_e{\bR_e}, {\bR_e}^T\mbs{\sigma}_e{\bR_e})$ for all rotation matrices ${\bR_e} \in SO(3)$, the group of proper orthogonal matrices in three dimensions. Thus, if a point $(\mbs{\epsilon}_e, \mbs{\sigma}_e)$ is in the local data set $E_e$, then so is the entire {\sl orbit} of the point by $SO(3)$.

Under these conditions, local optimality demands
\begin{equation}
\begin{split}
    &
    F_e(\mbs{\epsilon}_e,\mbs{\sigma}_e)
    = \\ &
    \min_{(\mbs{\epsilon}_e',\mbs{\sigma}_e') \in E_e}
    \min_{{\bR_e} \in SO(3)}
    \Big(
        W_e(\mbs{\epsilon}_e-{\bR_e}^T\mbs{\epsilon}_e'{\bR_e})
        +
        W_e^*(\mbs{\sigma}_e-{\bR_e}^T\mbs{\sigma}_e'{\bR_e})
    \Big) .
\end{split}
\end{equation}
The corresponding optimality condition is
\begin{equation}
\begin{split}
    &
    \frac{\partial W_e}{\partial\epsilon_{ij}}
    \frac{\partial}{\partial R_{mn}} (\epsilon'_{kl} R_{ki}R_{lj})
    +
    \frac{\partial W_e^*}{\partial\sigma_{ij}}
    \frac{\partial}{\partial R_{mn}} (\sigma'_{kl} R_{ki}R_{lj})
    - \\ &
    \frac{\partial}{\partial R_{mn}} (\Lambda_{ij} R_{ki}R_{kj})
    =
    0 ,
\end{split}
\end{equation}
where $\mbs{\Lambda} = \mbs{\Lambda}^T$ is a Lagrange multiplier enforcing the orthogonality of ${\bR_e}$. Evaluating the derivatives, we obtain, in matrix form,
\begin{equation}
    {\bR_e}^T
    \mbs{\epsilon}_e'
    {\bR_e}
    \left(\frac{\partial W_e}{\partial\mbs{\epsilon}_e}\right)
    +
    {\bR_e}^T
    \mbs{\sigma}_e'
    {\bR_e}
    \left(\frac{\partial W_e^*}{\partial\mbs{\sigma}_e}\right)
    =
    \mbs{\Lambda} .
\end{equation}
Transposing both sides and using tensor symmetry we obtain
\begin{equation}
    \left(\frac{\partial W_e}{\partial\mbs{\epsilon}_e}\right)
    {\bR_e}^T
    \mbs{\epsilon}_e'
    {\bR_e}
    +
    \left(\frac{\partial W_e^*}{\partial\mbs{\sigma}_e}\right)
    {\bR_e}^T
    \mbs{\sigma}_e'
    {\bR_e}
    =
    \mbs{\Lambda} ,
\end{equation}
whence it follows that
\begin{equation}\label{eq:b953h5}
\begin{split}
    &
    ({\bR_e}^T\mbs{\epsilon}_e'{\bR_e})
    \left(\frac{\partial W_e}{\partial\mbs{\epsilon}_e}\right)
    +
    ({\bR_e}^T\mbs{\sigma}_e'{\bR_e})
    \left(\frac{\partial W_e^*}{\partial\mbs{\sigma}_e}\right)
    = \\ &
    \left(\frac{\partial W_e}{\partial\mbs{\epsilon}_e}\right)
    ({\bR_e}^T\mbs{\epsilon}_e'{\bR_e})
    +
    \left(\frac{\partial W_e^*}{\partial\mbs{\sigma}_e}\right)
    ({\bR_e}^T\mbs{\sigma}_e'{\bR_e}) .
\end{split}
\end{equation}
These equations are now to be solved for the local optimal principal directions $\{\bR_e,\ e=1,\dots,m\}$, e.~g., by recourse to a Newton-Raphson iteration based on a convenient parametrization of $SO(3)$.

A simple situation arises when the local state $(\mbs{\epsilon}_e, \mbs{\sigma}_e\}$ is itself isotropic, i.~e., $\mbs{\epsilon}_e$ and  $\mbs{\sigma}_e\}$ have the same principal directions, and $W_e$ and $W_e^*$ are chosen to be isotropic. In this case, the optimality condition (\ref{eq:b953h5}) is satisfied if ${\bR_e}^T\mbs{\epsilon}_e'{\bR_e}$ and $DW_e$ and ${\bR_e}^T\mbs{\sigma}_e'{\bR_e}$ and $DW_e^*$ {\sl commute}, which in turn holds if and only if ${\bR_e}^T\mbs{\epsilon}_e'{\bR_e}$ and $\mbs{\epsilon}_e$ and ${\bR_e}^T\mbs{\sigma}_e'{\bR_e}$ and $\mbs{\sigma}_e$ have the same eigenvectors. Introducing the representations
\begin{equation}
\begin{split}
    &
    \mbs{\epsilon}_e= {\bQ_e}^T \be_e {\bQ_e} ,
    \qquad
    \mbs{\sigma}_e = {\bQ_e}^T \bs_e {\bQ_e} ,
    \\ &
    \mbs{\epsilon}'_e = {\bQ'_e}^T \be'_e {\bQ'_e} ,
    \qquad
    \mbs{\sigma}'_e = {\bQ'_e}^T \bs'_e {\bQ'_e} ,
\end{split}
\end{equation}
with ${\bQ_e}, {\bQ'_e} \in SO(3)$ and $\be_e$, $\bs_e$, $\be'_e$, $\bs'_e$ diagonal, local optimality then requires
\begin{equation}
    {\bR_e} = {\bQ'_e} {\bQ_e}^{-1},
\end{equation}
which determines explicitly the optimal data point in the $SO(3)$-orbit of $(\be'_e, \bs'_e)$.

In general, since the local states $(\mbs{\epsilon}_e, \mbs{\sigma}_e\}$ follow from independent Euler-Lagrange equations, eqs.~(\ref{eq:jp88tm}), they need not be exactly isotropic and the general optimality equations (\ref{eq:b953h5}) need to be solved in order to determine the optimal principal directions of the local data points.

\subsection{Numerical analysis of convergence}

\begin{figure} [H]
\centering
\mbox{
\subfigure[]{\includegraphics[width=0.475\linewidth]{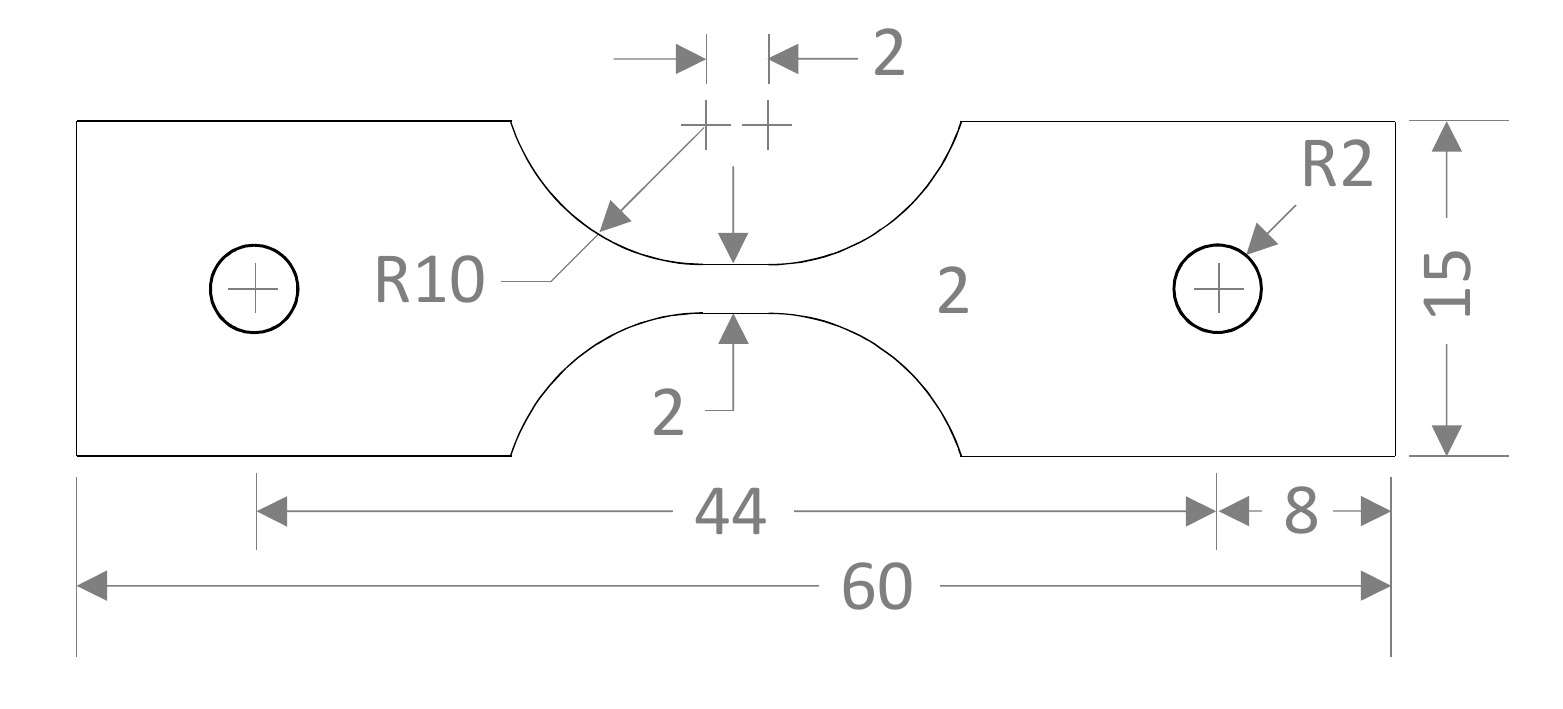}}
\subfigure[]{\includegraphics[width=0.425\linewidth]{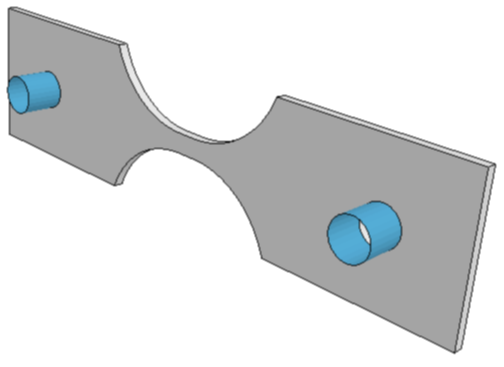}}}
\caption{a) Sketch of the simulation set-up of a thin tensile specimen loaded in tension \cite{Fen12}. The thickness of the sample is  $1\,\rm{mm}$ for the three dimensional model. b) Isometric view of the simulation set-up in 3D consisting of two rigid pins and the tensile specimen.}
\label{fig:thinstrip}
\end{figure}

\begin{figure} [H]
\centering
\mbox{
\subfigure[]{\includegraphics[width=0.475\linewidth]{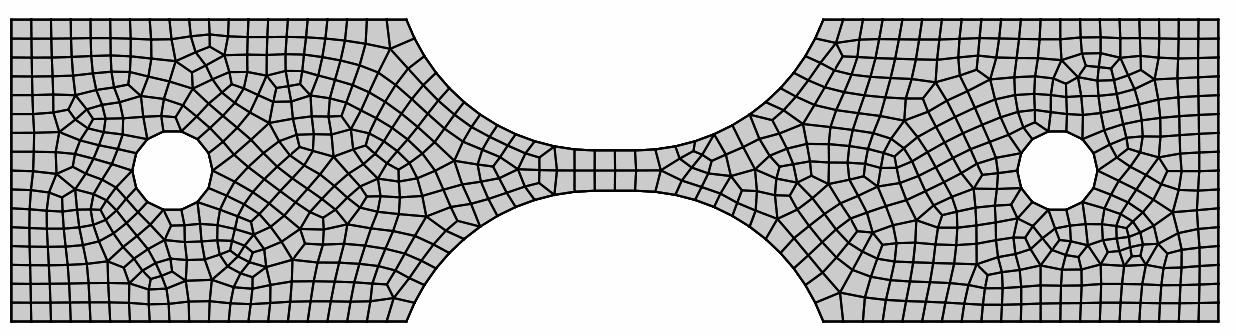}}
\subfigure[]{\includegraphics[width=0.475\linewidth]{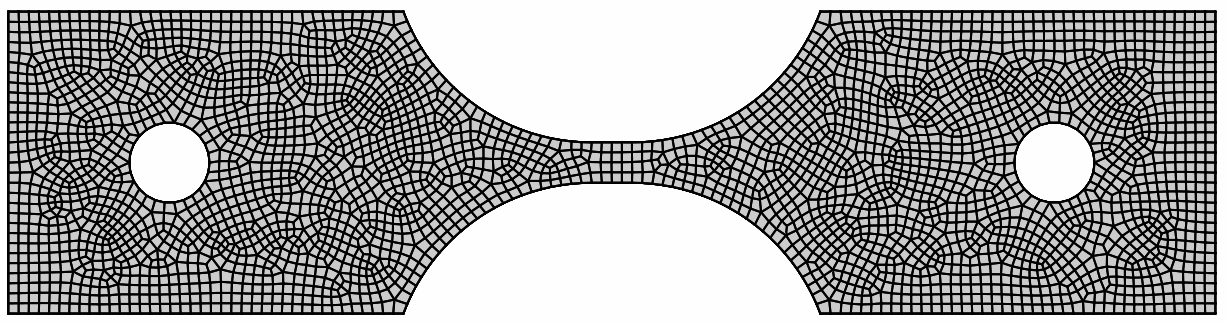}}}
\caption{a) Coarse mesh with 811 element and an average element edge length $h\approx1\rm{mm}$; b) Fine mesh with 6428 elements and an average element edge length $h=0.5\rm{mm}$.}
\label{fig:mesh}
\end{figure}

Similarly to the case of truss analysis considered earlier, we revisit the question of convergence of the linear-elasticity data-driven solver with respect to the data set. We specifically consider the problem of the thin tensile specimen shown in Fig.~\ref{fig:thinstrip}, cf.~\cite{Fen12}. The specimen is loaded by two rigid pins and contains a short gauge section undergoing ostensibly homogeneous deformation. By contrast, the regions surrounding the pin-loaded holes undergo complex heterogeneous deformations. Two finite-element discretizations are used in order to ascertain the influence of mesh resolution. The coarse mesh, Fig.~\ref{fig:mesh}a, consists of 811 elements and one element across the thickness, whereas the fine mesh, Fig.~\ref{fig:mesh}b, consists of 3,214 elements in two-dimensions and 6,428 elements in three-dimensions, respectively. These discretizations correspond to average element sizes of $h=1\rm{mm}$ for the coarse mesh and $h=0.5\rm{mm}$ for the fine mesh. The mesh consists of eight-node hexahedral elements containing eight Gauss quadrature points each.

Sampling requirements are reduced by virtue of the plane-stress conditions of the problem under consideration. Specifically, only a neighborhood of the subspace $\sigma_{13} = \sigma_{23} = \sigma_{33} = 0$ in stress space needs to be covered by the data. We accomplish this requirement by sampling an appropriate region of the $(\sigma_{11}, \sigma_{22}, \tau_{12})$ stress plane on a uniform cubic grid. The corresponding strains $(\epsilon_{11}, \epsilon_{22}, \epsilon_{12})$ then obey an isotropic linear-elastic law. A reference isotropic linear-elastic solid of the type (\ref{eq:ury933}), unrelated to the actual material behavior sampled by the material data, is used in the data-driven calculations.  These reductions effectively limit the material data set to a three-dimensional space.

\begin{figure} [H]
\centering
  \includegraphics[width=0.75\linewidth]{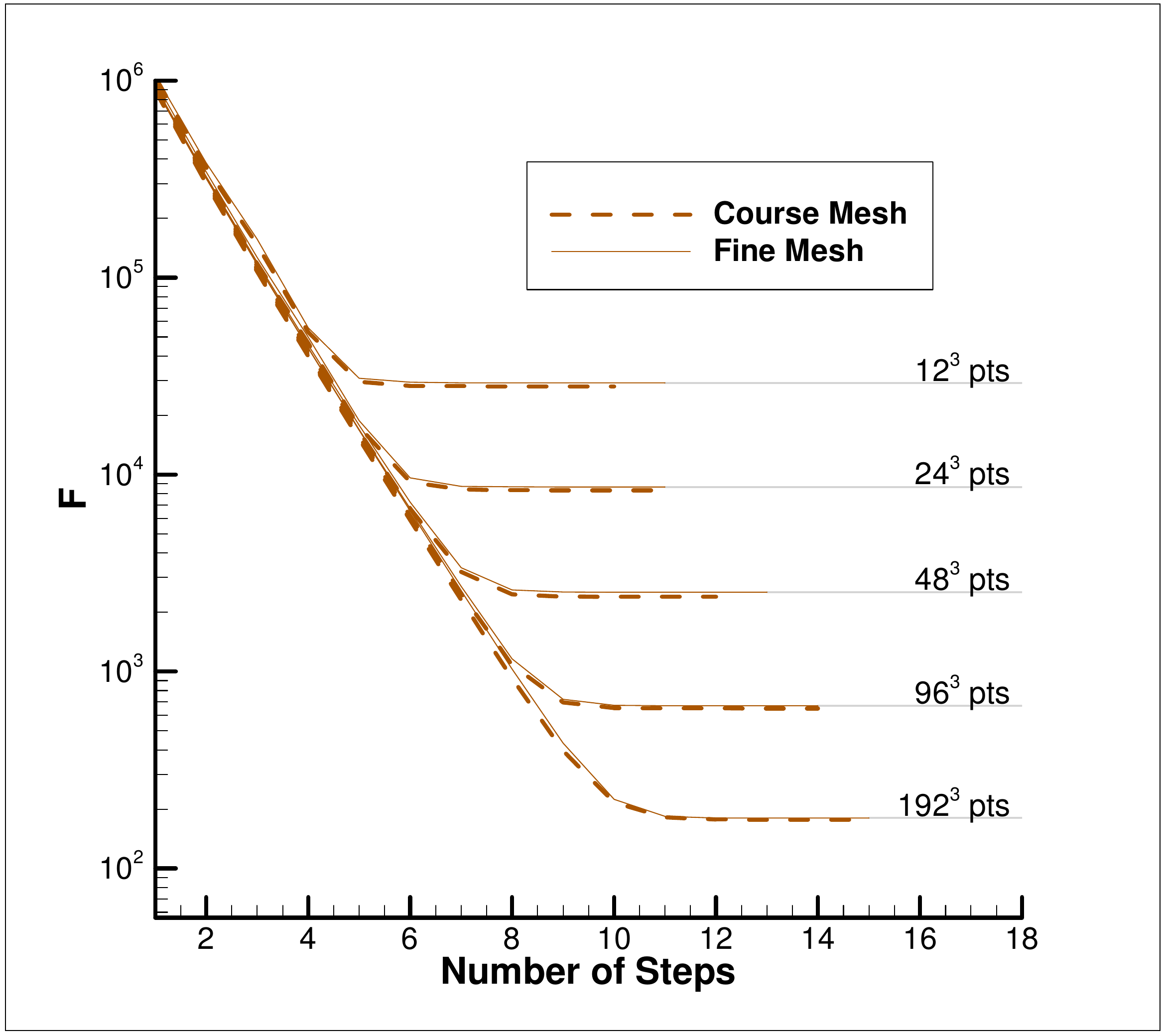}
\caption{Linear-elastic tensile specimen. Convergence of the local material-data assignment iteration. Functional $F$ decays through increasing data resolution in a three dimensional sampling of the plane stress space for both mesh resolutions.}
\label{fig:LE_FunctionalDecay}
\end{figure}

As in the case of the data-driven truss problem, we focus on the questions of convergence for a data-driven linear-elastic solver with respect to local data assignment, or step-wise convergence, and with respect to the data set.  In Fig.~\ref{fig:LE_FunctionalDecay}, the global functional $F$ is again shown to decay through iteration for both mesh resolutions on increasingly large data sets. The number of iterations to convergence increases with the material-data sample size but, remarkably, remains modest in all cases.

\begin{figure} [H]
\centering
  \includegraphics[width=0.75\linewidth]{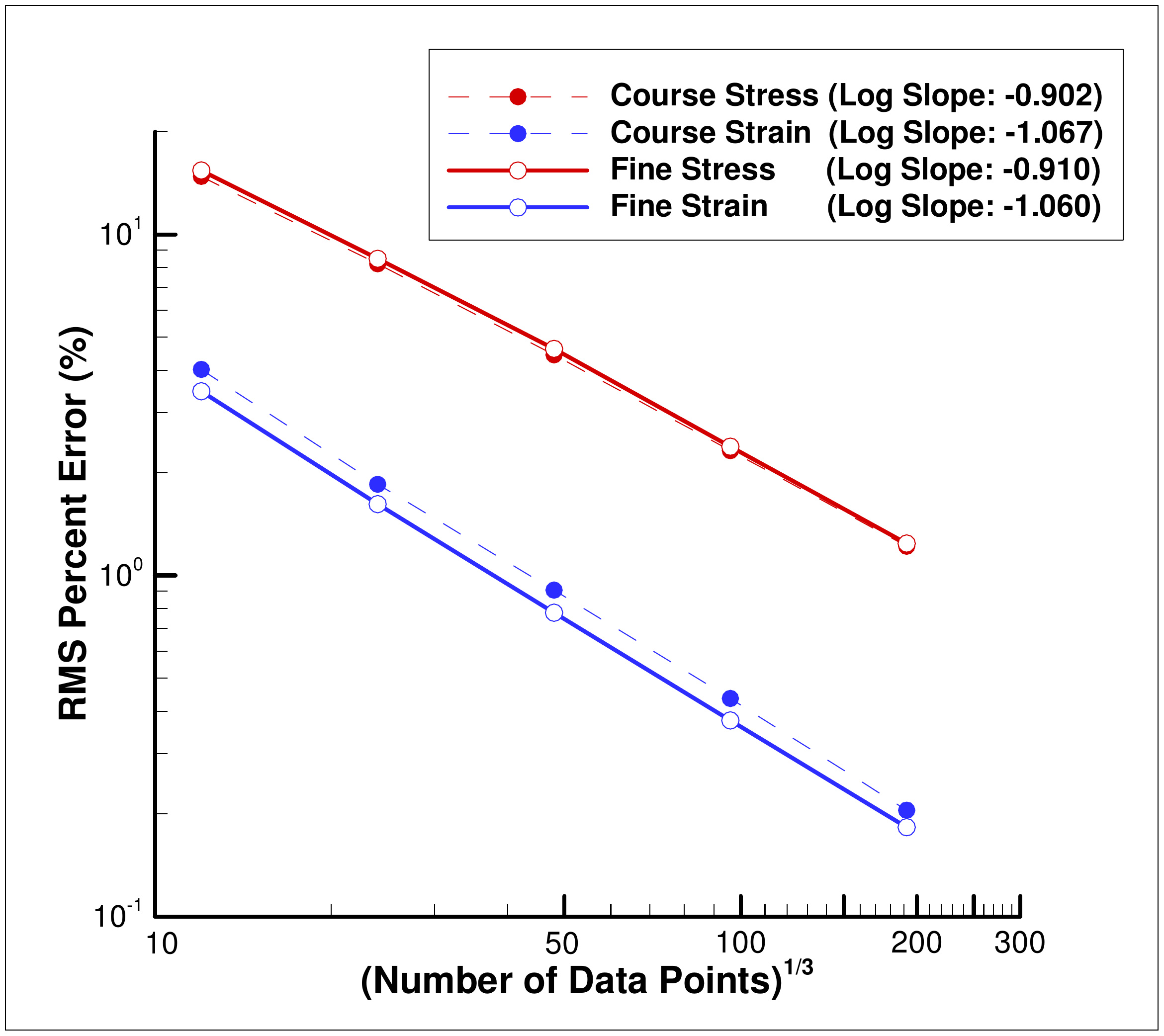}
 \caption{Linear-elastic tensile specimen. Convergence with respect to sample size. RMS errors decay linearly in data resolution for both stresses ($\mbs{\sigma}$) and strains ($\mbs{\epsilon}$).}
\label{fig:LE_ErrorDecay}
\end{figure}

To calculate percent error with respect to the reference solution we re-define the RMS error metric as \begin{subequations} \label{eq:LE_PercentError}
\begin{align}
    \sigma_{(\%RMS)}
    & =
    \left(
        \frac
        {
            \sum_{e=1}^m
            w_e W^*(\mbs{\sigma}_e-\mbs{\sigma}^{\text{ref}}_e)
        }
        {
            \sum_{e=1}^m
            w_e
            W^*(\mbs{\sigma}_e^{\text{ref}})
        }
    \right)^{1/2} ,
    \\
    \epsilon_{(\%RMS)}
    & =
    \left(
        \frac
        {
            \sum_{e=1}^m
            w_e
            W(\mbs{\epsilon}_e-\mbs{\epsilon}^{\text{ref}}_e)
        }
        {
            \sum_{e=1}^m
            w_e
            W(\mbs{\epsilon}_e^{\text{ref}})
        }
    \right)^{1/2} ,
\end{align}
\end{subequations}
where $W$ and $W^*$ are the strain and complementary-energy densities as calculated using the reference solution moduli, respectively. Plots of these errors against the cubic root of the number of data points are shown in Fig.~\ref{fig:LE_ErrorDecay}. The plots are indicative of ostensibly linear convergence, in keeping with the analytical estimates derived next.

%% file: Analysis.tex
We proceed to abstract from the preceding examples a general class of data-driven problems and to establish some of their fundamental properties by way of analysis. We consider systems whose state is characterized by points in a certain phase space $Z$. For instance, in the case of linear elasticity, the system of interest is an elastic solid occupying a certain domain $\Omega$ and the state of the system is defined by the pair $(\mbs{\epsilon}(\bx), \mbs{\sigma}(\bx))$, where $\mbs{\epsilon}(\bx)$ is the strain field and $\mbs{\sigma}(\bx)$ is the stress field, both defined over $\Omega$. In this case, the phase space $Z$ of the elastic solid is an appropriate space of pairs $(\mbs{\epsilon}(\bx), \mbs{\sigma}(\bx))$ of strain and stress fields over $\Omega$.

We particularly wish to characterize the states of the system that are in a constraint set $C$ of states satisfying essential constraints and conservation laws. For instance, in the running example of linear elasticity we may wish to determine states $(\mbs{\epsilon}(\bx), \mbs{\sigma}(\bx)) \in Z$ satisfying compatibility, i.~e., such that
\begin{subequations}
\begin{align}
    &
    \mbs{\epsilon}(\bx)
    =
    \nicefrac{1}{2} \big( \nabla\bu(\bx) + \nabla\bu^T(\bx) \big) ,
    &
    \bx \in \Omega ,
    \\ &
    \bu(\bx) = \bar{\bu}(\bx),
    &
    \bx \in \partial\Omega_D ,
\end{align}
\end{subequations}
for some displacement field $\bu$ over $\Omega$ and prescribed displacements $\bar{\bu}$ over the Dirichlet boundary $\partial\Omega_D$, and satisfying equilibrium, i.~e., such that
\begin{subequations}
\begin{align}
    &
    \nabla\cdot\mbs{\sigma}(\bx)
    +
    \fb(\bx)
    =
    \mbs{0} ,
    &
    \bx \in \Omega , \\
    &
    \mbs{\sigma}(\bx) \bn(\bx) = \bar{\bt}(\bx),
    &
    \bx \in \partial\Omega_N ,
\end{align}
\end{subequations}
for some applied body force field $\fb$ over $\Omega$ and tractions $\bar{\bt}$ over the Neumann boundary $\partial\Omega_N$, with unit outer normal $\bn$.

Classically, the problem is closed by putting forth a material law restricting the set of admissible states to a graph $E$ in $Z$. For instance, in linearized elasticity the material law may classically take the form a nonlinear Hooke's law
\begin{equation}
    \mbs{\sigma}(\bx) = DW\big(\mbs{\epsilon}(\bx)\big),
    \qquad
    \bx \in \Omega ,
\end{equation}
where $W$ is the strain-energy density of the material. The set $E$ then consists of the set of strain and stress fields satisfying the material law at all material points in $\Omega$. The classical solution set is then the intersection $E \cap C$, consisting of states of the system satisfying the essential constraints, the conservation laws and the material law simultaneously. In the case of linear elasticity, the classical solutions would consist of compatible strain fields and equilibrium stress fields satisfying the material law at all material points. In general, the cardinality of the solution set $E \cap C$ depends on the transversality of $C$ with respect to $E$, depending on which transversality, the solution set may be empty or non-empty, in which latter case the solution set may consist of a single point, corresponding to uniqueness of the solution, or multiple points.

\begin{figure} [H]
\centering
\includegraphics[width=0.65\linewidth]{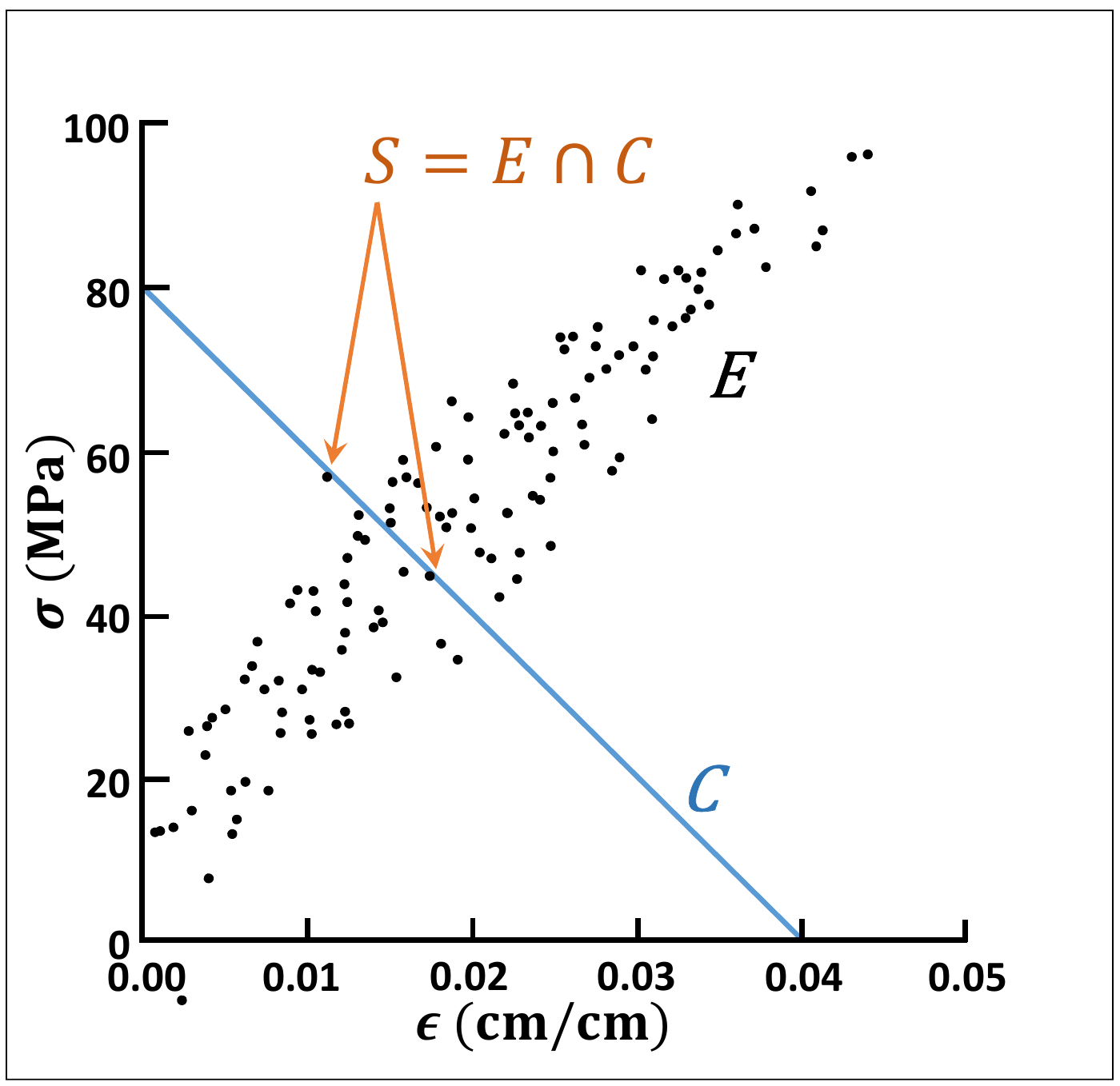}
\caption{Schematic of a local material set $E$ consisting of a finite number of states obtained, e.~g., from experimental testing. Also shown is a possible constraint set $C$ and near intersections between $E$ and $C$.}
\label{fig:Anaysis:Fig1}
\end{figure}

In contrast to the classical problem just formulated, here we suppose that the material response is not known exactly and, instead, it is imperfectly characterized by a set, also denoted $E$, consisting locally at every material point, e.~g., of a finite collection of states obtained by means of experimental testing, cf.~Fig.~\ref{fig:Anaysis:Fig1}. Under such conditions, $E \cap C$ is likely to be empty even in cases when solutions could reasonably be expected to exist. It is, therefore, necessary to replace the overly-rigid problem of determining $E \cap C$ by a suitable {\sl relaxation} thereof. To this end, we begin by introducing a norm $| \cdot |$ in phase space. For instance, for truss structures we may choose
\begin{equation}\label{eq:hyu1y5}
    | x |^2
    =
    \sum_{e=1}^m
    w_e
    \left(
        C_e \varepsilon_e^2
        +
        \frac{\sigma_e^2}{C_e}
    \right) ,
\end{equation}
with $x \equiv (\epsilon_e,\sigma_e)_{e=1}^m$ denoting a generic point in phase space $Z$. For discretized linear-elastic solids we may choose
\begin{equation}\label{eq:qh2jrz}
    | x |^2
    =
    \sum_{e=1}^m
    w_e
    \left(
        \mathbb{C}_e\mbs{\epsilon}_e \cdot \mbs{\epsilon}_e
        +
        \mathbb{C}_e^{-1}\mbs{\sigma}_e\cdot\mbs{\sigma}_e
    \right) ,
\end{equation}
with $e$ labeling the integration points in the discretization and $x \equiv (\mbs{\epsilon}_e,\mbs{\sigma}_e)_{e=1}^m$ denoting a generic point in phase space $Z$. Finally, for continuum linear elasticity we may choose
\begin{equation}
    | x |^2
    =
    \int_\Omega
    \left(
        \mathbb{C}(\bx)\mbs{\epsilon}(\bx) \cdot \mbs{\epsilon}(\bx)
        +
        \mathbb{C}^{-1}(\bx) \mbs{\sigma}(\bx) \cdot\mbs{\sigma}(\bx)
    \right)
    \, dx .
\end{equation}
with $x \equiv \{\big( \mbs{\epsilon}(\bx), \mbs{\sigma}(\bx) \big),\ \bx \in \Omega\}$ denoting a generic point in phase space $Z$.

Based on this metrization of phase space, we define the {\sl data-driven problem} as the double minimum problem
\begin{equation}\label{eq:9cbrn4}
    \min_{y \in E} \min_{x \in C} | y - x |
    =
    \min_{y \in E} {\rm dist}(y, C),
\end{equation}
or, equivalently,
\begin{equation}\label{eq:9cbrn5}
    \min_{x \in C} \min_{y \in E} | x - y |
    =
    \min_{x \in C} {\rm dist}(x, E) .
\end{equation}
Thus, the aim of the data-driven problem, as expressed in (\ref{eq:9cbrn4}), is to {\sl find the point in the material-data set that is closest to satisfying the essential constraints and conservation laws}, or, as expressed in (\ref{eq:9cbrn5}), to {\sl find the point in the constraint set that is closest to the material-data set}. In the particular example of linear elasticity, the aim of the data-driven problem, as expressed in (\ref{eq:9cbrn4}), is to find the point in the material-data set that is closest to being compatible and in equilibrium, or, as expressed in (\ref{eq:9cbrn5}), to find the compatible equilibrium point that is closest to the material-data set.

We note that the data-driven problems considered in Sections~\ref{sec:Truss} and Sections~\ref{sec:LE} are indeed examples of (\ref{eq:9cbrn4}) and (\ref{eq:9cbrn5}) with norms (\ref{eq:hyu1y5}) and (\ref{eq:qh2jrz}), respectively.

\subsection{Finite-dimensional case: Convergence with respect to sample size}

We begin by considering systems whose local states take values in a finite-dimensional phase space $Z$. The global state of the system is then characterized by a point $x \in Z$. The essential constraints and conservation laws pertaining to the system have the effect of constraining its global state to lie on a subset $C$ of $Z$. For instance, for linear-elastic trusses such as considered in the preceding section, the local phase space $Z_e$ of bar $e$ is the space of pairs $(\epsilon,\sigma)$, where $\epsilon$ is axial strain of a bar and $\sigma$ the corresponding axial stress. The global phase space of the entire truss is then $Z = Z_1 \times \cdots \times Z_m$, where $m$ is the number of bars in the truss. In addition, the constraint set $C$ is the affine space of compatible and equilibrated states of stress and strain in the truss.

The data-driven problem (\ref{eq:9cbrn4}) is now formulated by specifying a set $E$ of possible material states in $Z$. For instance, in the case of a truss a local material set $E_e$ of the form shown in Fig.~\ref{fig:Anaysis:Fig1} may be supplied for every bar $e$ of the truss and the global material set is then $E = E_1 \times \cdots \times E_m$. We note that, if $E$ is compact, e.~g., consisting of a finite collection of points, then the corresponding data-driven problem has solutions by the Weierstrass extreme-value theorem.

We proceed to consider the question of convergence with respect to the data set. Specifically, we suppose that a sequence $(E_k)$ of data sets is supplied that approximates increasingly closely a limiting data set $E$. The particular case in which $E$ is a graph concerns convergence of data-driven solutions to classical solutions. For instance, the approximations $(E_k)$ may be the result of an increasing number of experimental tests sampling the behavior of a material characterized by a---possibly unknown---stress-strain curve $E$. The sequence of approximate material data sets $(E_k)$ generates in turn a sequence of approximate data-driven problems
\begin{equation}
    \min_{x_k \in C} \min_{y_k \in E_k} | x_k - y_k | ,
\end{equation}
and attendant approximate solutions $(x_k)$. We wish to ascertain conditions under which $(x_k)$ converges to solutions of the $E$-problem.

Conditions ensuring such convergence at a well-defined convergence rate are given in the following proposition. Henceforth, we denote by ${\rm dist}(x,E)$ the distance from a point $x \in Z$ to a subset $E \subset Z$, i.~e.,
\begin{equation}
    {\rm dist}(x,E) = \inf_{y \in E} | x - y | ,
\end{equation}
and by $P_Y x$ the projection of $x \in Z$ onto a subspace $Y$ of $Z$, i.~e.,
\begin{equation}
   | x - P_Y x | = {\rm dist}(x, Y) .
\end{equation}

\begin{proposition}\label{prop:conv}
Let $(E_k)$ be a sequence of finite subsets of $Z$, $E$ a subset of $Z$ and $C$ a subspace of $Z$. Let $x$ be an isolated point of $E \cap C$ and let $x_k, y_k \in Z$ be such that
\begin{equation}
    (x_k,y_k)
    \in
    {\rm argmin}\{ |x-y|,\ x \in C,\ y \in E_k \} .
\end{equation}
Suppose that:
\begin{itemize}
\item[i)] There is a sequence $\rho_k \downarrow 0$ such that
\begin{equation}
    {\rm dist}(z, E_k) \leq \rho_k, \qquad \forall z \in E .
\end{equation}
\item[ii)] There is a sequence $t_k \downarrow 0$ such that
\begin{equation}
    {\rm dist}(z_k, E) \leq t_k, \qquad \forall z_k \in E_k .
\end{equation}
\item[iii)] (Transversality) There is a constant $0 \leq \lambda < 1$ and a neighborhood $U$ of $x$ such that
\begin{equation}
    | P_C z - x | \leq \lambda | z - x | ,
\end{equation}
for all $z \in E \cap U$.
\end{itemize}
Then,
\begin{equation}\label{eq:prop:rate1}
    | x_k - x |
    \leq
    \frac{t_k + \lambda ( t_k + \rho_k )}{1 - \lambda} ,
\end{equation}
and, therefore, $\lim_{k\to\infty} | x_k - x | = 0$.
\end{proposition}
\begin{proof}
By assumption (i), we can find $z_k \in E_k$ such that
\begin{equation}
    | z_k - x | \leq \rho_k .
\end{equation}
By optimality,
\begin{equation}
    {\rm dist}(y_k,C)
    \leq
    {\rm dist}(z_k,C) .
\end{equation}
Then, we have
\begin{equation}
    | x_k - y_k |
    =
    {\rm dist}(y_k,C)
    \leq
    {\rm dist}(z_k,C)
    \leq
    | z_k - x |
    \leq
    \rho_k .
\end{equation}
By assumption (ii), we can find $z_k \in E$ such that
\begin{equation}
    | y_k - z_k | \leq t_k .
\end{equation}
By the triangular inequality, we have
\begin{equation}
    | x_k - x |
    \leq
    | x_k - P_C z_k | + | P_C z_k - x | .
\end{equation}
By the contractivity of projections,
\begin{equation}
    | x_k - P_C z_k |
    =
    | P_C y_k - P_C z_k |
    =
    | P_C (y_k - z_k) |
    \leq
    | y_k - z_k |
    \leq
    t_k .
\end{equation}
In addition, by transversality, we have
\begin{equation}
    | P_C z_k - x |
    \leq
    \lambda
    | z_k - x | ,
\end{equation}
with $0 \leq \lambda < 1$. Triangulating again,
\begin{equation}
    | z_k - x |
    \leq
    | z_k - y_k | + | y_k - x_k | + | x_k - x | .
\end{equation}
Collecting all the preceding estimates, we obtain
\begin{equation}
    | x_k - x |
    \leq
    t_k + \lambda ( t_k + \rho_k + | x_k - x | )
\end{equation}
whence (\ref{eq:prop:rate1}) follows.
\hfill$\square$
\end{proof}

\begin{figure} [H]
\centering
\includegraphics[width=0.75\linewidth]{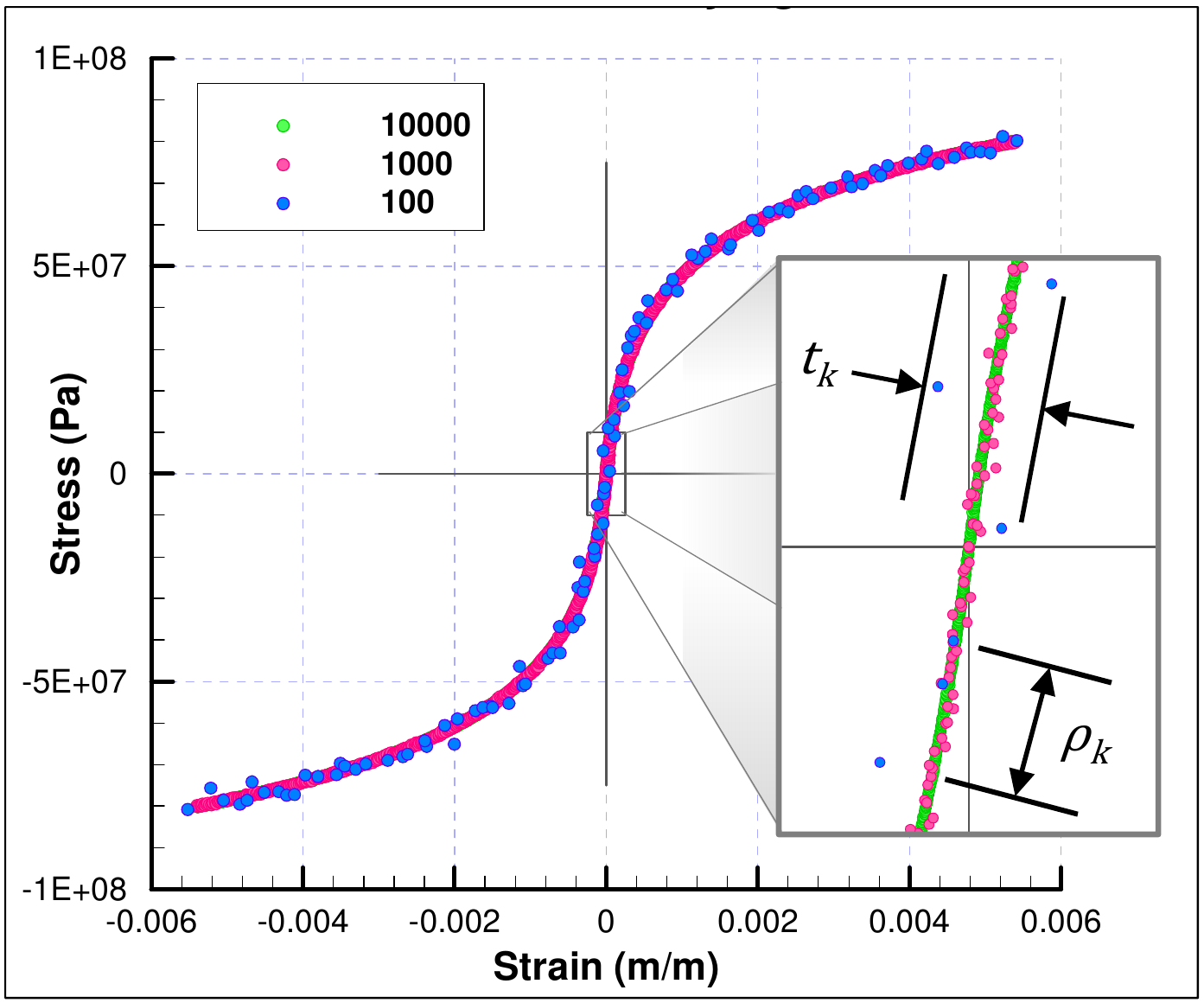}
\caption{Schematic of convergent sequence of material-data sets. The parameter $t_k$ controls the spread of the material-data sets away from the limiting data set and the parameter $\rho_k$ controls the density of material-data point.}
\label{fig:Anaysis:Fig2}
\end{figure}

The preceding proposition presumes that the limiting data set $E$ and the constraint subspace $C$ are transversal at isolated intersections, which are identified with the limiting solutions. For instance, $E$ may be a Lipschitz continuous graph in a neighborhood of classical solutions with $E$ not contained in $C$ in that neighborhood. In the particular case in which the limiting material response is linear, the transversality condition reduces to the requirement that the displacement stiffness matrix of the system be non-singular. Assumptions (i) and (ii) set how the sequence of data sets $E_k$ must approximate $E$. Thus, (i) ensures that there are approximate data points increasingly and uniformly closer to any point of $E$, whereas (ii) ensures that there are no outliers in the approximate data sets such as could spoil the approximate solutions, cf.~Fig.~\ref{fig:Anaysis:Fig2}. In particular, (ii) requires $E_k$ to be contained uniformly within the $t_k$-neighborhood of $E$.

Precise convergence rates with respect to, e.~g., the number of sampling points $N_k = \# \, E_k$, are derived from the preceding theorem if the sequences $\rho_k$ and $t_k$ are related to $N_k$. In particular, we have the following,

\begin{corollary}\label{cor:Analysis}
Assume that there are constants $C_1 > 0$, $C_2 > 0$ and $\alpha > 0$ such that $\rho_k \leq C_1 N_k^{-\alpha}$ and $t_k \leq C_2 N_k^{-\alpha}$. Then
\begin{equation}
    | x_k - x |
    \leq
    \frac{C_2 + \lambda ( C_1 + C_2 )}{1 - \lambda} N_k^{-\alpha} .
\end{equation}
\end{corollary}

The numerical convergence rates of Sections~\ref{sec:Truss} and \ref{sec:LE} indeed conform to this estimate. Thus, for the case of truss structures with noise-free data, $\rho_k$ and $t_k$ scale as $N_k^{-1}$, resulting in a linear convergence rate $\alpha = 1$, cf.~Fig.~\ref{fig:NoNoiseConverge}, whereas for noisy data $\rho_k$ and $t_k$ scale as $N_k^{-1/2}$, resulting in a linear convergence rate $\alpha = 1/2$, cf.~Fig.~\ref{fig:NoiseConverge}. For the case of plane-stress linear elasticity with noise-free data, $\rho_k$ and $t_k$ scale as $N_k^{-3}$, resulting in a linear convergence rate $\alpha = 3$, cf.~Fig.~\ref{fig:LE_ErrorDecay}.

\subsection{Infinite-dimensional case: Convergence with respect to mesh size}

The linear-elastic case considered in Section~\ref{sec:LE} differs from the truss case of Section~\ref{sec:Truss} in that it obtained by discretization of an infinite-dimensional problem. The question then naturally arises of convergence of the data-driven problem with respect to the mesh size. Consider, for simplicity, a sequence of discretizations of the domain into constant strain triangles of size $h_k$. Let $x \equiv (\mbs{\epsilon}, \mbs{\sigma})$ be the classical solution and $x_{h_k} \equiv (\mbs{\epsilon}_{h_k}, \mbs{\sigma}_{h_k})$ the corresponding sequence of finite-element solutions. Simultaneously consider a sequence $(E_k)$ of local material-data sets satisfying conditions (i) and (ii) of Prop.~\ref{prop:conv} for some sequences $\rho_k \downarrow 0$ and $t_k \downarrow 0$. Additionally suppose that the sequence of discretizations is regular in the sense that
\begin{equation}
    | x_{h_k} - x |
    \leq
    C h_k
\end{equation}
for some constant $C > 0$ and that the transversality constants $\lambda_k$ of the sequence of finite-element models does not degenerate, i.~e.,
\begin{equation}
    0 \leq \lambda_k \leq \lambda ,
\end{equation}
for some $\lambda < 1$. Then, by Prop.~\ref{prop:conv} we have
\begin{equation}
    | x_{k,h_k} - x |
    \leq
    | x_{k,h_k} - x_{h_k} |
    +
    | x_{h_k} - x |
    \leq
    \frac{t_k + \lambda ( t_k + \rho_k )}{1 - \lambda}
    +
    C h_k ,
\end{equation}
where $x_{k,h_k} \equiv (\mbs{\epsilon}_{k,h_k}, \mbs{\sigma}_{k,h_k})$ denotes the data-driven solution corresponding to the $E_k$ material-data set and the $h_k$ discretization. It thus follows that if $\rho_k$ and $t_k$ are controlled by the mesh size, i.~e., there is a constant $C > 0$ such that
\begin{equation}
    \rho_k < C h_k, \qquad t_k < C h_k ,
\end{equation}
then
\begin{equation}
    | x_{k,h_k} - x |
    \leq
    C h_k ,
\end{equation}
for some constant $C > 0$ not renamed. We thus conclude that the data-driven paradigm is robust with respect to spatial discretization, in the sense that it preserves convergence provided that the fidelity of the data set increases appropriately with increasing mesh resolution.

%% file: Summary.tex
We have formulated a new computing paradigm, which we refer to as {\sl data-driven computing}, consisting of formulating calculations {\sl directly} from experimental material data and pertinent essential constraints and conservation laws, thus bypassing the empirical material modeling step of conventional computing altogether. The data-driven solver specifically seeks to assign to each material point of the computational model the closest possible state from a prespecified material-data set, while simultaneously satisfying the essential constraints and conservation laws. Optimality of the local state assignment is understood in terms of a figure of merit that penalizes distance to the data set in phase space. The resulting data-driven problem thus consists of the minimization of a distance function to the data set in phase space subject to constraints set forth by the essential constraints and conservation laws.

We have investigated the performance of the data-driven solver with the aid of two particular examples of application, namely, the static equilibrium of nonlinear three-dimensional trusses and of finite-element discretized linear-elastic solids. In these cases, the penalty function in phase space may be regarded as representing a linear-comparison solid with an initial state of strain and stress. The equilibrium constraint can be conveniently enforced by means of Lagrange multipliers. The corresponding stationarity equations correspond to the solution of two linear-static equilibrium problems for the comparison solid. We have formulated a local data assignment algorithm by which each member of the truss is pegged to a particular point in the data set. The algorithm terminates when the local state of every member of the truss is in the Voronoi cell of its assigned data point in phase space. We show, by way of numerical testing, that the data-driven solver possesses good convergence properties both with respect to the number of data points and with regards to the local data assignment iteration.

The variational structure of the data-driven problem confers robustness to the solver and renders it amenable to analysis. By exploiting this connection, we show that data-driven solutions converge to classical solutions when the data set approximates a limiting constitutive law with increasing fidelity. By virtue of this property, we may regard data-driven problems as a generalization of classical problems in which the material behavior is defined by means of an arbitrary data set in phase space, not necessarily a graph. In particular, classical solutions are recovered precisely when the data set coincides with the graph of a material law.

Whereas the data-driven paradigm has been formulated in the context of computational mechanics and, specifically elastic quasistatic problems, we believe that its range and scope is much larger. Indeed, field theories governed by linear or nonlinear elliptic partial-differential equations should be amenable, upon discretization, to an analogous treatment. Extensions to dynamical problems are also straightforward. Indeed, dynamics essentially adds inertia forces in the equations of motion that are independent---and do not affect the description---of the material behavior. By contrast, inelastic materials raise the fundamental problem of sampling history-dependent material behavior. Such sampling should provide appropriate coverage of possible processes and evolutions of the system and is thus likely to result in exceedingly large and complex data sets. The use of tools from Data Science and Big Data management may be expected to be particularly beneficial in dealing with such data sets.

We close by pointing out that the traditional computing paradigm has insulated problems from the data on which their solution is based. Removing this barrier creates a powerful new tool in the arsenal of scientific computing. With data-driven computing, data sets can be used directly to provide predictive analysis capability for unmodeled materials. Traceability and inherent measures of data fidelity enable both deeper investigations into data-solution relationships and natural alerts for appropriate model use. Having the ability to tie solution results back to specific data points within a set allows for the creation of a new kind of causality in material analysis. The data-driven paradigm can also ensure the collection of descriptive data sets for prospective uses. Error measures highlight data regions that require additional resolution, as well as point the analyst toward sensitivities within the solution-source relations. These methods can be used to check if a constitutive relation based on a certain data-set is capable of performing a desired simulation {\sl prior to the analysis}. Tying the solution back to the data set also establishes an elegant way of limiting model accuracy to the resolution of the source data.  Additionally, it should be noted that material models have specific regimes over which they are developed. However, the models themselves are easily used outside this development range.  Especially with regards to empirical {\sl ad-hoc} curve fits, such overreach can neither be justified nor easily prevented. By directly using a data set in calculations, attempts to simulate beyond the data regime are met and penalized by large calculated errors, regardless of how the user receives the data set. These tangible and intangible benefits add considerable appeal to data-driven solvers beyond their mere usefulness as numerical schemes.